\documentclass[12pt]{article}
\usepackage[utf8]{inputenc}

\usepackage{amsmath}
\usepackage{amssymb}
\usepackage{amsthm}
\usepackage{color}
\usepackage{bm}
\usepackage[colorlinks,linkcolor=blue,anchorcolor=green,citecolor=red]{hyperref}
\usepackage{appendix}
\usepackage{graphicx}
\usepackage{subfigure}
\usepackage{booktabs}
\usepackage{epstopdf}
 \usepackage{caption}
\usepackage{subfigure}
\allowdisplaybreaks[4]
\usepackage[utf8]{inputenc}          
\usepackage[T1]{fontenc}  

\usepackage{url}
\usepackage{hyperref}
\usepackage{float}
\title{ \textcolor{black}{Coupled Chemical Reactions: Effects of Electric Field, Diffusion and Boundary Control}}

	\bigskip
   \author{Shixin Xu\thanks{Zu Chongzhi Center for Mathematics and Computational Sciences, Duke Kunshan University, 8 Duke Ave, Kunshan, Jiangsu, China.}  
	\and Robert  Eisenberg \thanks{Department of Applied Mathematics, Illinois Institute of Technology, Chicago, IL, 60616, USA; Department of Physiology and Biophysics, Rush University, Chicago, IL, 60612, USA.}
	\and Zilong Song \thanks{ Math and Statistics Department, Utah State University,   Old Main Hill
		Logan, UT 84322.}
	\and  Huaxiong Huang\thanks{Corresponding author, Research Center for Mathematics, Advanced Institute of Natural Sciences, Beijing Normal University, Zhuhai, Guangdong, 519088, China; Guangdong Provincial Key Laboratory of Interdisciplinary Research and Application for Data Science, BNU-HKBU United International College, Zhuhai, Guangdong, 519088, China; Laboratory of Mathematics and Complex Systems, MOE, Beijing Normal University, 100875 Beijing, China;  Department of Mathematics and Statistics York University, Toronto, ON, M3J 1P3, Canada (hhuang@uic.edu.cn). }}
\date{}
\usepackage{amsmath}
\usepackage{geometry}
\newcommand\bD{\boldsymbol{D}}
\newcommand\bE{\boldsymbol{E}}
\newcommand\bj{\boldsymbol{j}}

\newtheorem{rmk}{Remark}[section]
\providecommand{\keywords}[1]
{
  \small	
  \textbf{\textit{Keywords---}} #1
}
\begin{document}

\maketitle
\begin{abstract}
	Chemical reactions involve the movement of charges, and this  \textcolor{black}{work} presents a mathematical model for describing chemical reactions in electrolytes. The model is developed using an energy variational method that aligns with classical thermodynamics principles. It encompasses both electrostatics and chemical reactions within consistently defined energetic and dissipative functionals. Furthermore, the energy variation method is extended to account for open systems that involve the input and output of charge and mass. Such open systems have the capability to convert one form of input energy into another form of output energy.
In particular, a two-domain model is developed to study a reaction system with self-regulation and internal switching, which plays a vital role in the electron transport chain of mitochondria responsible for ATP generation—a crucial process for sustaining life. Simulations are conducted to explore the influence of electric potential on reaction rates and switching dynamics within the two-domain system.  \textcolor{black}{ It shows that the electric potential inhibits the oxidation reaction while accelerating
the reduction reaction.}
\end{abstract}
\keywords{ Open system, Mass action, Electrochemistry }
\section{Introduction}

Theories for chemical reactions based on the law of mass actions often ignore electric effects~\cite{brush1976kind,garber1995maxwell}, even though the reactants, catalysts and enzymes of chemistry and biology depend on charge interactions for much of their function~\cite{simpson1998maxwell,zangwill2013modern,chree1908mathematical}. Recently, mathematical models have been developed in the electrochemical tradition~\cite{fraggedakis2021theory,bazant2013theory,bazant2005current,van2010diffuse,bonnefont2001analysis}, some using variational princlies~\cite{wang2020field,wang2022some}. 
Here, we extend this work by proposing a general thermodynamics-consistent framework for electrochemical reactions both in the bulk and on the interface in biological systems. 
Specifically, we investigate the electric effects on the rate of chemical reactions using several reduced (`toy') models that show interesting  dynamics and interactions between diffusion, reaction, and electric fields of ionic solutions.

Various consistent frameworks have been developed for nonequilibrium systems  based on the second law of thermodynamics and Onsager's linear response theory. Onsager's variational principle, which was first proposed in~\cite{onsager1931reciprocal1,onsager1931reciprocal2} and later generalized by Edwards \cite{edwards1974theory}, Doi \cite{doi1983variational}, and Qian \cite{qian2006variational}, is a popular approach to irreversible systems. It is based on the maximum dissipation principle proposed by Lord Rayleigh \cite{strutt1871some} motivated by the analysis of uncharged systems.
The principle states that for a system described by variables $\alpha_1,\cdots,\alpha_n$ that describe the displacement from thermodynamic equilibrium with the (corresponding) rates $\dot{\alpha}_1,\cdots,\dot{\alpha}_n$, and with free energy $F(\alpha_1,\cdots,\alpha_n)$, the thermodynamic flux $\dot{\alpha}_i$ follows the dynamic path that minimizes the Rayleigh function. This function is the sum of the dissipation function $\Phi$ and the rate of the change of the free energy $\dot{F}$ in an isothermal system. \cite{wang2021onsager} provide a more detailed review. In order to include the fluid kinetic energy \cite{finlayson1972existence} for fluid equations,    Wang et al. \cite{wang2021generalized, yang2016hydrodynamic} proposed the Generalized Onsager principle (GOP) for both reversible and irreversible processes.  

Energetic Variational Approach (EnVarA) is another powerful tool proposed by Liu et al.~\cite{ryham2006energetic,eisenberg2010energy,xu2014energetic}.  The general framework of EnVarA is a combination of statistical physics and nonlinear thermodynamics. All the physics are integrated into the definitions of total energy $E$ and dissipation functional $\Delta$. 
The Least action principle  \cite{feynman2006feynman}  yields the conservative force $F_{con}$ by taking the variation of the action functional with respect to the flow map in the Lagrange frame of reference. The Maximum principle \cite{onsager1931reciprocal1,onsager1931reciprocal2} yields dissipative force $F_{dis}$ by taking the variation of the rate function. The final equation of momentum is achieved by balancing these two forces balance $F_{con} = F_{dis}$. A more detailed review can be found in \cite{wang2022some}.

Based on the consideration of the second law of thermodynamics, in particular, the requirement that the rate of energy dissipation needs to be non-positive, Ren et al. \cite{ren2010continuum,zhao2021thermodynamically,zhang2014derivation} proposed a more straightforward method built on a concept in \cite{gennes2004capillarity}. First, the energy functional $E$ is defined according to different physical fields.   Based on the laws of conservation, the kinematic assumptions are listed with unknown terms like fluxes and stresses. By taking the time derivative of the total energy $\frac{dE}{dt}$, those unknowns could be described so each term in $\frac{dE}{dt}$ is negative.  Following  Ren's Method, Shen et al. proposed an energy variation method \cite{shen2020energy,shen2022energy} with predefined energetic and dissipative functionals. The unknown terms in the kinematics assumptions  are obtained by matching $\frac{dE}{dt}$ with a predefined dissipative function $\Delta$.

 \textcolor{black}{In recent years, the variational method has been instrumental in proposing various dynamic boundary conditions that consider the possible short-range interactions of materials with solid walls. For instance, Qian et al. \cite{qian2006variational,liu2012hydrodynamic} utilized the Onsager principle to establish the general Navier boundary condition (GNBC), which captures the dynamics of contact lines. Furthermore, building upon EnVarA, Liu et al. \cite{liu2019energetic} introduced a new category of dynamic boundary conditions, further generalized in \cite{knopf2021phase}. Most of the existing research has primarily focused on closed systems with no exchanges through the boundary, subsequently incorporating flux conditions as part of the boundary dissipation \cite{qian2006variational}.  However, it's important to note that a majority of biological systems and all engineered devices are open systems characterized by the flow of flux both into and out of the considered domain.
Addressing this, Brunet et al. \cite{brunet2004generalized} were the pioneers in considering the mass flux effect using the Onsager variational method in linear electrohydrodynamic responses. Building on this, Xu et al. \cite{xu2019generalized} extended the concept by establishing the generalized Lorentz reciprocal theorem for complex fluids and non-isothermal systems. This paper introduces a thermodynamically consistent framework tailored for open systems. The proposed framework incorporates both energetic and dissipative components at the boundary and within the bulk, while also considering chemical reactions.
}

 \textcolor{black}{The law of mass action \cite{nelson2008lehninger} is a foundational principle in chemistry that establishes a relationship between the concentrations of reactants and products in an ideal chemical reaction and its equilibrium constant. In recent times, various models have been developed to describe chemical reactions based on the principles of thermodynamics \cite{ge2016mesoscopic,yong2012conservation,mielke2017non}. Wang et al. \cite{wang2020field,wang2022some,liu2021structure} extended this framework with the EnVarA approach to encompass chemical reaction systems described by rate constants. 
Our methods extend the pioneering work of Wang et al. \cite{wang2020field,wang2022some}, who explored reactions not involving charges or electrodynamics. 
This generalization enables us to investigate the influence of electric potential on chemical reaction rates often described by the Butler-Volmer equation \cite{butler1924studies,erdey1930theorie,dickinson2020butler}. The generalization equips us with the necessary tools for modeling biochemical reactions.}


The rest of the paper is organized as follows. In Section \ref{section:model}, we derive the general   field theory for open system with flux on the boundary  and then the method is use to derive equations for an ionic system with reaction  in bulk region, on the boundary and on the interior interface. The simulation results for a bidomain rection system are presented in Section \ref{section:simulation}. The discussions  and conclusions are shown in Section \ref{section:discussion}.

\section{Mathematical Models }\label{section:model}
For a closed system, the First Law of Thermodynamics states that the rate of change of the sum of  the kinetic energy $\mathcal{K}$ and  the
internal energy $\mathcal{U}$ is equal to the sum of the rates of change of work $\mathcal{W}$ and heat $\mathcal{Q}$, so
$\frac {d(\mathcal{K}+\mathcal{U})}{dt}=\frac{d\mathcal{W}}{dt}+\frac{d\mathcal{Q}}{dt}.$
 \textcolor{black}{In the context of standard statistical physics, the internal energy $\mathcal{U}$ considers the interactions among particles. These interactions can be categorized as either short range, such as hardcore interactions, or long range, such as Coulomb electrostatic interactions.}
The Second Law of Thermodynamics, in the isothermal case,  is given by,
$T\frac{d\mathcal{S}}{dt}=\frac{d\mathcal{Q}}{dt}+\Delta,$
where $T$ is temperature, $\mathcal{S}$ is entropy and $\Delta\geq0$ is entropy production.  As a reformulation of the linear response assumption, this entropy production functional can be represented as the sum of various rates such as the velocities and the strain rates.  By subtracting the Second Law from the First Law, under the isothermal assumption, assuming both laws are valid for nonequilibrium open systems, we have,
\begin{equation}
	\frac{dE}{dt}=\frac{d\mathcal{W}}{dt}-\Delta,      
\end{equation}
where $E=\mathcal{K} +\mathcal{F}$ is  the total
energy and $\mathcal{F} := \mathcal{U} -\mathcal{TS}$ is  the Helmholtz free energy.In case no external forces or fields are applied, i.e. $\frac{d\mathcal{W}}{dt} = 0$, we can derive the dissipation law $\frac{dE}{dt} = -\Delta$.

Open systems have some fluxes that flow in or out through the boundary and have distinctive energy dissipation laws that differ from those of closed systems.  \textcolor{black}{They have the distinct inputs and outputs that characterize devices in engineering. }  In open systems, energy can change both because of the flux across the boundary and also because of the change in dissipation.
For open systems, we assume the following energy law
\begin{equation}\label{energydispplawwithsupp}
	\frac{dE}{dt}=\mathcal{P}_{E,\partial\Omega} -\Delta,    
\end{equation} 
where $\mathcal{P}_{E,\partial\Omega}$ is the energy exchange rate through energy due to the variation of total particles, energy. It is important to note that both the total energy and dissipation functionals incorporate contributions from both the bulk region and the boundary.

\subsection{Bulk reaction}

We consider  a domain that has reactions in bulk and connects to an infinite large reservoir with fixed chemical potential on the boundary.  
We focus on  elementary reactions containing $N$ species  \textcolor{black}{$\{X_1^{z_1},X_2^{z_2},\cdots, X_N^{z_N}\}$}
 \textcolor{black}{\begin{equation}\label{reacton_de}
	a_1 X_1^{z_1}+a_2 X_2^{z_2}+\cdots+a_N X_N^{z_N}\overset{k_f}{\underset{k_r}{\rightleftharpoons}}  b_1 X_1^{z_1}+b_2 X_2^{z_2}+\cdots+b_N X_N^{z_N},
\end{equation}}
where $k_f$ and $k_r$ are the forward and backward reaction rates,   \textcolor{black}{ $C_i$  is  the concentration  of $X_i$, respectively. }Here  $a_i$  and $b_i$ are the stoichiometric coefficients, and $z_i$ is the valence of $i^{th}$ species.   If let $\gamma_i = b_i-a_i$. Charge conservation implies $\sum_{i=1}^N \gamma_i z_i=0$.

Based on the  \textcolor{black}{law of conservation of mass  \cite{de2019traite}} and Maxwell equations, we have the following kinematic  assumptions  
\begin{equation}\label{assumption_de}
	\left\{
	\begin{array}{l}
		\frac{\partial C_i}{\partial t} =-\nabla\cdot \boldsymbol{j}_i +\gamma_i\ \mathcal{R},\\[3mm]
		\nabla\cdot(\bD) = \sum_{i=1}^Nz_iC_iF,\\[3mm]
		\nabla\times \bE = \boldsymbol{0},
	\end{array}
	\right.
\end{equation}
where  $\boldsymbol{j}_i, i=1,\cdots, N$ are the passive fluxes, $z_i$ is the valence of $i_{th}$ particle, $F$ is Faraday constant and $\mathcal{R} = \mathcal{R}_f -\mathcal{R}_r$ is the net reaction rate function with forward and reverse reactions rate $\mathcal{R}_f$  and  $\mathcal{R}_r$.  $\bD$ is Maxwell's electrical displacement field and $\bD = \varepsilon_0\varepsilon_r \bE$ with electric field $\bE$, dielectric constant $\varepsilon_0$ and relative dielectric constant $\varepsilon_r$. 
The equation $ \nabla\times \bE = \boldsymbol{0}$
implies that there exists  an  \textcolor{black}{electric potential $\phi$} such that $\bE = -\nabla \phi$.  


The boundary conditions are 
\begin{equation}\label{bd_flux}
	\left\{
	\begin{array}{ll}
		\boldsymbol{j}_i\cdot\boldsymbol{n} = j_{i,ex},\quad i=1\cdots, N, & \mbox{on~} \partial\Omega,\\
		\boldsymbol{D}\cdot\boldsymbol{n} = 0, & \mbox{on~} \partial\Omega.
	\end{array}\right.
\end{equation}
where $j_{i,ex}$ is the flux of $i_{th}$ ion supplied from an external source and to be determined later with fixed reservoir chemical potential, $\bm{n}$ is the unit outward normal vector. 
\begin{rmk}
	By multiplying $ \textcolor{black}{z_iF}$ on both sides of the first  equation, we have 
	\begin{eqnarray}
		\frac{\partial}{\partial t}(\nabla \cdot\bD) =\sum_{i=1}^Nz_iF\frac{\partial C_i}{\partial t}
		= -\sum_{i=1}^N\nabla\cdot(z_iF\bj_i)+ \sum_{i=1}^N \textcolor{black}{z_iF}\gamma_i\mathcal{R}
		=-\sum_{i=1}^N\nabla\cdot(z_iF\bj_i)
	\end{eqnarray}
	which is consistent with the electrostatic Maxwell equations. Treatment of transient problems, involving displacement currents is needed to deal with some important experimental work \cite{belevich2007exploring,bloch2004catalytic,verkhovsky2006elementary,blomberg2012mechanism,cai2018network,verkhovskaya2008real}. 
\end{rmk}

The total energetic functional  is defined as the summation of entropies of mixing, 
internal energy and electrical static energy \cite{truesdell1969rational,wang2020field},  
\begin{eqnarray}\label{totalenergy}
	E &=& E_{ent}+E_{int}+E_{ele}\nonumber\\
	&= &\sum_{i=1}^N \int_{\Omega}RT\left\{ C_i\left(\ln{\left(\frac{C_i}{c_0}\right)}-1 \right)\right\} dx+\int_{\Omega} \sum_{i=1}^NC_iU_idx 
	+
	\int_{\Omega} \frac{\boldsymbol{D}\cdot\boldsymbol{E}}{2}dx,
\end{eqnarray}
where $R$ is universal gas constant, $T$ is temperature, $U_i$ is the standard free energy density of $i_{th}$ particle (temperature of 298 K; concentration $1 M$) \cite{nelson2008lehninger}. 

Then the  \textcolor{black}{electrochemical} potentials (used in the following derivations) are defined by
\begin{equation}
	\tilde\mu_i= \mu_i +z_i\phi F = RT\ln\frac{C_i}{c_0} +U_i +z_i\phi F, ~i=1,\cdots, N.
\end{equation}

It is assumed in the present work that the dissipation of the system energy is due to passive diffusion and chemical reaction. Additional dissipations (and energies for that matter) can be included if needed in later applications.

\begin{eqnarray}
	\Delta 
	&=& \int_{\Omega}\left\{\sum_{j=1}^N\frac{RT}{D_iC_i}|\bj_i|^2+  RT\mathcal{R}\ln \left(\frac{\mathcal{R}_f}{\mathcal{R}_r}\right)\right\} dx+\sum_{i=1}^N\int_{\partial\Omega}\frac{1}{g_i}|j_{i,ex}|^2dS,
\end{eqnarray}
where $\mathcal{R}=\mathcal{R}_f-\mathcal{R}_r$  and $g_i$ is the conductivity  of $i_{th}$ species on the boundary. 

The boundary energy communication rate $\mathcal{P}_{E,\partial\Omega}$ is induced by the mass communication flux with energy density  \textcolor{black}{$\tilde{\mu}_{i,ex}$} 
\begin{equation}
	\mathcal{P}_{E,\partial\Omega} = -\sum_{i=1}^N\int_{\partial\Omega} \tilde{\mu}_{i,ex} j_{i,ex} dS,
\end{equation}
where $\tilde{\mu}_{i,ex}$ is the fixed chemical potential  of $i_{th}$ species  $\tilde{\mu}_i$ in the  reservoir.

 \textcolor{black}{Taking the time derivative of the energy functional  \eqref{totalenergy} yields}

 \textcolor{black}{	\begin{eqnarray}\label{eqn: dedtingeneral}
\frac{dE}{dt}
	&=&\int_{\Omega}\sum_{i=1}^N\left\{\nabla\tilde{\mu}_i\cdot \boldsymbol{j}_i\right\}dx +\int_{\Omega}\mathcal{R} \sum_{i=1}^N\gamma_i  \tilde{\mu}_i dx-\int_{\partial\Omega} \sum_{i=1}^N (\tilde{\mu}_i-\tilde{\mu}_{i,ex}) \bj_i\cdot\boldsymbol{n} dS -\int_{\partial\Omega} \tilde{\mu}_{i,ex} \bj_i\cdot\boldsymbol{n} dS.\nonumber \\
	&=&-\Delta +\mathcal{P}_{E,\partial\Omega}.
\end{eqnarray}}
 \textcolor{black}{where the detailed derivation can be found in the Appendix.}

By comparing with the dissipation function, we have 
\begin{equation}\label{bulkfit}
	\left\{
	\begin{array}{ll}
		\bm{j}_i = -\frac{D_i}{RT}C_i\nabla\tilde\mu_i,~ i=1,\cdots, N    &  \mbox{in~} \Omega \\
		\bm{j}_i\cdot\bm{n} = g_i(\tilde{\mu}_i-\tilde{\mu}_{i,ex})    & \mbox{on~} \partial\Omega\\
		RT\ln \left(\frac{\mathcal{R}_f}{\mathcal{R}_r }\right) =  - \sum_{i=1}^N\gamma_i  \tilde{\mu}_i=  RT \ln \left(\frac{\Pi^N_{i=1} \left(\frac{C_{i}}{c_0}\right)^{a_i}}{\Pi^N_{i=1} \left(\frac{C_{i}}{c_0}\right)^{b_i}} e^{\frac{\Delta U}{RT}}\right),&  \mbox{on~} \partial\Omega.
	\end{array}
	\right.
\end{equation}
where   $A=-\sum_{i=1}^N\gamma_i \tilde{\mu}_i = \sum_{i=1}^N(a_i-b_i) \tilde{\mu}_i$is the  local affinity  \cite{de2019reaction} and the last equation is  \textcolor{black}{part of the De Donder-Weyl theory }\cite{bazant2013theory,sekimoto2010stochastic}.

If we assume  the reaction rate function is 
\begin{equation}
	\mathcal{R}=\mathcal{R}_f-\mathcal{R}_r =k_{f,0}\Pi_{i=1}^N\left(\frac{C_i}{c_0}\right)^{a_i}-k_{r,0}\Pi_{i=1}^N\left(\frac{C_i}{c_0}\right)^{b_i}\nonumber
\end{equation}
where   $k_{f,0}$ and $k_{r,0}$ are  forward and backward reaction rate constants with unit $\frac{M}{s} = \frac{Mol}{Ls} $ \cite{ozcan2022equilibrium},
then  \textcolor{black}{De Donder-Weyl theory}   yields 
\begin{equation}
	\frac{k_{r,0}}{k_{f,0}} =k_{eq}= e^{\frac{-\Delta U}{RT}},
\end{equation}
where $\Delta U = \sum_{i=1}^N(a_i  -b_i) U_i$ and $k_{eq}$ is the equilibrium constant \cite{ozcan2022equilibrium}.

The governing equations for the bulk reaction are given as
\begin{equation}\label{model_ele}
\left\{
\begin{array}{l}
\frac{\partial C_i}{\partial t} =\nabla\cdot (D_i \nabla C_i+D_i\frac{z_iF}{RT}C_i\nabla\phi) + \textcolor{black}{\gamma_i \left(k_{f,0}\Pi_{i=1}^N\left(\frac{C_i}{c_0}\right)^{a_i}-k_{r,0}\Pi_{i=1}^N\left(\frac{C_i}{c_0}\right)^{b_i}\right)},\\[3mm]
-\nabla\cdot(\varepsilon_0\varepsilon_r \nabla \phi) = \sum_{i=1}^Nz_iFC_i,
\end{array}
\right.
\end{equation}
with
 boundary conditions 
\begin{equation}\label{bd_flux}
\left\{
\begin{array}{ll}
\boldsymbol{j}_i\cdot\boldsymbol{n} =  g_i(\tilde{\mu}_i-\tilde{\mu}_{i,ex}), i=1\cdots N, & \mbox{on~} \partial\Omega,\\
\boldsymbol{D}\cdot\boldsymbol{n} = 0, & \mbox{on~} \partial\Omega.
\end{array}\right.
\end{equation}
The energy law of this system could be written as follows 
\begin{eqnarray}
\frac{dE}{dt} &=&-\int_{\Omega}\left\{\sum_{j=1}^N\frac{D_iC_i}{RT}|\nabla\tilde{\mu}_i|^2+  RT\mathcal{R}\ln \left(\frac{\mathcal{R}_f}{\mathcal{R}_r}\right)\right\} dx -\sum_{i=1}^N\int_{\partial\Omega}g_i|\tilde{\mu}_{i}-\tilde{\mu}_{i,ex}|^2dS\nonumber\\
&&-\sum_{i=1}^N\int_{\partial\Omega}g_i\tilde{\mu}_{i,ex}(\tilde{\mu}_{i}-\tilde{\mu}_{i,ex})dS\nonumber\\
&=& -\Delta +\mathcal{P}_{E,\partial\Omega}.
\end{eqnarray}
Here the boundary condition of concentration is Robin type  \textcolor{black}{that allows flux across a boundary `resistance'.}  
\begin{itemize}
\item If the boundary conductivity is zero $g_i=0$, then we obtain the nonflux boundary condition $\bm{j}_i\cdot\bm{n}=0$  \textcolor{black}{that describes an insulating boundary and a closed system.} The corresponding energy law is 
\begin{eqnarray}
\frac{dE}{dt} &=&-\int_{\Omega}\left\{\sum_{j=1}^N\frac{D_iC_i}{RT}|\nabla\tilde{\mu}_i|^2+  RT\mathcal{R}\ln \left(\frac{\mathcal{R}_f}{\mathcal{R}_r}\right)\right\} dx = -\Delta.
\end{eqnarray}
with zero boundary energy communication rate $\mathcal{P}_{E,\partial\Omega} = 0$.
\item If the conductance is infinitely large, the Dirichlet boundary condition is achieved $\tilde\mu_i=\tilde{\mu}_{i,ex}$ on the boundary.  \textcolor{black}{This Dirichlet condition is called a 'voltage clamp' in biophysics. Note the voltage clamp is an open system requiring the injection of current and energy provided by an external system usually called a voltage clamp amplifier.} 
\item If the conductance is a function of chemical potential,  \textcolor{black}{as may be the case in some membrane proteins, including channels and transporters,} i.e. $g_i=\frac{j_{i,0}}{\tilde{\mu}_i-\tilde{\mu}_{i,ex}}$, where $j_{i,0}$ is constant, then we obtain the Neumann boundary condition $\bm{j}_i\cdot\bm{n} = j_{i,0}$. In this case, the corresponding energy law is 
\begin{eqnarray}
\frac{dE}{dt} &=&-\int_{\Omega}\left\{\sum_{j=1}^N\frac{D_iC_i}{RT}|\nabla\tilde{\mu}_i|^2+  RT\mathcal{R}\ln \left(\frac{\mathcal{R}_f}{\mathcal{R}_r}\right)\right\} dx -\sum_{i=1}^N\int_{\partial\Omega}\tilde{\mu}_{i} j_{i,0}dS.
\end{eqnarray}
\end{itemize}
 \textcolor{black}{Our approach is able to handle a wide range of boundary conditions. For example, it can include complex current, voltage, and concentration relations needed to describe some ion channels and transporters.}

\subsection{ \textcolor{black}{Chemical reactions on the boundary}}
 \textcolor{black}{We now extend our treatment to include chemical reactions on the boundary, for example, as might happen on the } surface of a  \textcolor{black}{metal} electrode 

 \textcolor{black}{\begin{equation}\label{electro_reaction}
a_1 X_1^{z_1}+a_2 X_2^{z_2}+\cdots+a_N X_N^{z_N}+\Delta z e^{-1}\overset{k_f}{\underset{k_r}{\rightleftharpoons}}  b_1 X_1^{z_1}+b_2 X_2^{z_2}+\cdots+b_N X_N^{z_N},
\end{equation}}
 \textcolor{black}{where electrons are supplied through surface  $\Gamma$ with   the electrode potential $\phi_p$ and $\sum_{i=1}^N\gamma_iz_i+\Delta z =0$.    
For example, if $X_1, X_2,\cdots X_k$ are oxidized species and $X_{k+1},\cdots X_N$ are reduced species, then $a_{k+1}=a_{k+2}=\cdots=a_{N}=0$ and $b_1=b_2=\cdots=b_k=0$, i.e. 
 \textcolor{black}{\begin{equation}\label{electro_reaction}
a_1 X_1^{z_1}+a_2 X_2^{z_2}+\cdots+a_{k} X_k^{z_k}+\Delta z e^{-1}\overset{k_f}{\underset{k_r}{\rightleftharpoons}}  b_{k+1} X_{k+1}^{z_{k+1}}+b_{k+2} X_{k+2}^{z_{k+2}}+\cdots+b_N X_N^{z_N},
\end{equation}}}

\begin{figure}[h]
\centering
\includegraphics[width=.25\textwidth]{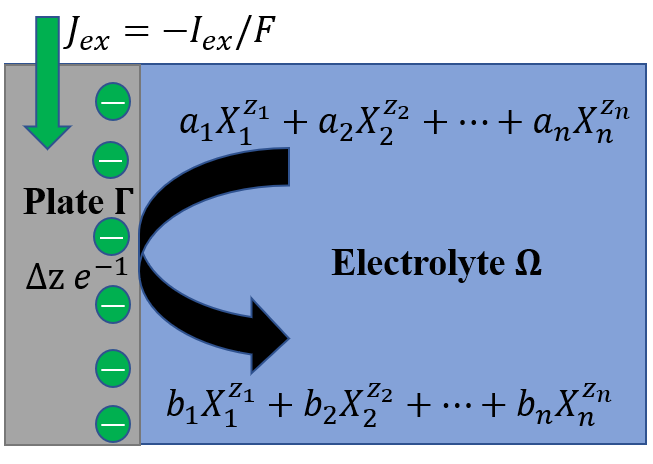}
\caption{ Schematic of reaction on the  \textcolor{black}{metal electrode} with external electron textcolor{blue}{input} $J_{ex} = \frac{-I_{ex}}{F}$. }
\label{fig:electrodereaction}
\end{figure}

During the derivation,  \textcolor{black}{we only focus on situations where the oxidized state exists only in solution near the surface of the metal plate \cite{bazant2012theory} as shown in Fig. \ref{fig:electrodereaction}. }

The conservation law yields the following kinematic assumptions, 
\begin{equation}\label{model_ele}
\left\{
\begin{array}{ll}
\frac{\partial C_i}{\partial t} =-\nabla\cdot\boldsymbol{j}_i, & \mbox{in}~ \Omega \\[3mm]
-\nabla\cdot(\varepsilon_0\varepsilon_r \nabla \phi) = \sum_{i=1}^Nz_iFC_i, & \mbox{in}~ \Omega
\end{array}
\right.
\end{equation}
with boundary conditions 
\begin{equation}\label{bd_flux}
\left\{
\begin{array}{ll}
\boldsymbol{j}_i\cdot\boldsymbol{n} = -\gamma_i \mathcal{R}, i=1\cdots N,  & \mbox{on~} \Gamma,\\
F\frac{\partial C_e}{\partial t} =  \mathbb{C}_p\frac{\partial (\phi-\phi_p)}{\partial t}= j_{ex}F- \Delta zF\mathcal{R}, &\mbox{on~} \Gamma\\
\boldsymbol{j}_i\cdot\boldsymbol{n}=0,  \phi = 0, &\mbox{on~}\partial\Omega/\Gamma,
\end{array}\right.
\end{equation}
where $C_e$ is the density of electrons on the surface of the plate, $\phi_p$ is the electric potential on the plate, and $\mathbb{C}_p$ is the capacitance,  $I_{ex}$ is the current supplied by an external amplifier and the corresponding inlet electron flux is $j_{ex}=-\frac{I_{ex}}{F}$.  Here we   use the fact that $FC_e = \mathbb{C}_p (\phi-\phi_p)$. 

Then the total energy is defined as the sum of mix energy and internal energy of ions, electric static energy in the bulk region, and the energy on the interface induced by electrons and  capacitor effect 
\begin{eqnarray}\label{eletro_totalenergy}
E &=& E_{ent}+E_{int}+E_{ele}+E_{\Gamma}\nonumber\\
&= &\underbrace{\sum_{i=1}^N \int_{\Omega}\left\{ RTC_i\left(\ln{\left(\frac{C_i}{c_0}\right)}-1 \right) +C_iU_i\right\}dx 
+
\int_{\Omega} \frac{\boldsymbol{D}\cdot\boldsymbol{E}}{2}dx}_{bulk}\nonumber\\
&&
+\underbrace{\int_{\Gamma} \left\{ RTC_e\left(\ln{\left(\frac{C_e}{c_{e,0}}\right)}-1 \right) +C_eU_e\right\}dx +\frac 1 2 \int_{\Gamma} \mathbb{C}_p(\phi-\phi_p)^2 dS}_{boundary}.
\end{eqnarray}
where $U_i$ and $U_e$ is the internal energy of $i_{th}$ ion  and electron, respectively.   \textcolor{black}{Here $c_{e,0}$ is the characteristic density.}

Then the corresponding  \textcolor{black}{electrochemical} potentials of ions and electrons are 
\begin{eqnarray}
\tilde{\mu}_{i}= \mu_i+z_iF\phi = U_i + RT \ln{\frac{C_i}{c_0}} +z_iF\phi,   \\
 \textcolor{black}{\tilde{\mu}_{e} = \mu_e-F\phi_p = U_e + RT \ln{\frac{C_e}{c_{e,0}}} -F\phi_p.} 
\end{eqnarray}

The dissipation function is defined as

\begin{eqnarray}\label{eletro_dissipation}
\Delta& =&\int_{\Omega} \sum_{j=1}^N\frac{D_iC_i}{RT}|\nabla\tilde\mu_i|^2dx+ \int_{\Gamma} RT\mathcal{R}\ln \left(\frac{\mathcal{R}_f}{\mathcal{R}_r}\right)  dS +\int_{\Gamma} \frac{ g}{F^2} (\tilde{\mu}_e-\tilde{\mu}_{ex})^2dS,
\end{eqnarray}
where we used the boundary reaction rate function $\mathcal{R} = \mathcal{R}_f-\mathcal{R}_r$, $\tilde{\mu}_{ex}$ is the external electron chemical potential of the reservoir that connects to the boundary. 

We analyze a setup (see Fig. 1) with a given current of electrons $I_{ex}$ applied to the plate $\Gamma$. This is a flux (really current) boundary condition, of the Neumann type in mathematical language.
 \textcolor{black}{The other boundaries of the system are controlled in a different way. They are insulators that help isolate the system so it communicates with the outside world in a limited well-defined way. On these insulating boundaries, no ions enter or leave the system on those other boundaries, i.e.} $\boldsymbol{J}_i\cdot \boldsymbol{n} = 0$, and the electric potential is fixed as $\phi_{ref}$. The current flow of electrons is only on the left   and the boundary energy power functional is defined as follows 
\begin{equation}
\mathcal{P}_{E,\partial\Omega} =  
\int_{\Gamma} \tilde{\mu}_e j_{ex}dx .
\end{equation}

\begin{rmk}
Integrating the first equation and using the boundary condition on $i_{th}$ ion yields  the law of mass conservation 
\begin{equation}
\frac{d}{dt}\int_{\Omega}C_idt = \int_{\Gamma} \gamma_i \mathcal{R} dS. 
\end{equation}
\end{rmk}

To apply the dissipation theorem, we need the derivative of energy with respect to time. 

 \textcolor{black}{\begin{eqnarray}\label{eqn:dedtboundary}
\frac{dE}{dt}
&=& \int_{\Omega}\sum_{i=1}^N \left\{ \nabla\tilde{\mu}_i\cdot \boldsymbol{j}_i\right\} dx-\int_{\Gamma}(-\sum_i^N\tilde{\mu}_i \gamma_i+\tilde{\mu}_e \Delta z)\mathcal{R}dS -\int_{\Gamma}\phi\frac{\partial  }{\partial t}\left(\boldsymbol{D}\cdot\boldsymbol{n} - FC_e\right)dS \nonumber\\
&&+\int_{\Gamma}(\tilde{\mu}_e-\tilde{\mu}_{ex}) j_{ex}dS +\int_{\Gamma}\tilde{\mu}_{ex} J_{ex}dS-\int_{\partial\Omega/\Gamma}\phi\frac{\partial \boldsymbol{D}\cdot\boldsymbol{n}}{\partial t}dS-\int_{\partial\Omega/\Gamma}\sum_i^N\tilde{\mu}_i \boldsymbol{j}_i\cdot\boldsymbol{n}dS\nonumber\\
&=&-\Delta+P_{E,\partial\Omega},
\end{eqnarray}}
 \textcolor{black}{where the detailed derivation is presented in the Appendix. }

To implement the dissipation principle, we now compare \textcolor{black}{ Eq. \eqref{eqn:dedtboundary}
   with the dissipation function defined in Eq. \eqref{eletro_dissipation}.} We are dealing with an open system so we include the dissipation (and energy) associated with the boundaries as well as the interior of the system.

\begin{equation}\label{singlematching}
\left\{
\begin{array}{ll}
\bj_i = -\frac{D_i}{RT}C_i\nabla\tilde\mu_i,~ i=1,\cdots, N, & \mbox{in} \Omega\\     
\boldsymbol{D}\cdot\boldsymbol{n} = FC_e   & \mbox{on}~ \Gamma\\
j_{ex} = \frac{g}{F^2}(\tilde{\mu}_{ex}-\tilde{\mu}_{e}),& \mbox{on}~ \Gamma\\
RT \ln\left(\frac{\mathcal{R}_f}{\mathcal{R}_r}\right) = -\sum_i^N \gamma_i\tilde{\mu}_i +\tilde{\mu}_e\Delta z,& \mbox{on}~ \Gamma\\
\boldsymbol{j}_i\cdot\boldsymbol{n} = 0,   \phi = 0,&\mbox{on}~\partial\Omega/\Gamma.
\end{array}
\right.
\end{equation}

The third  equation in \eqref{singlematching} yields 
\begin{equation}
\mathcal{R} = \mathcal{R}_f(1 - e^{\frac{-A_e}{RT}}). 
\end{equation}
with the affinity  
\begin{eqnarray}
 \textcolor{black}{A_e = -\sum_{i=1}^N \gamma_i \tilde{\mu}_i+\Delta z \tilde{\mu}_e 
= RT\ln\left(\left(\frac{C_e}{c_{e,0}}\right)^{\Delta z} \frac{\Pi^N_{i=1} \left(\frac{C_i}{c_0}\right)^{a_i}}{\Pi^N_{i=1} \left(\frac{C_i}{c_0}\right)^{b_i} k_{eq}  }e^{\frac{ \Delta zF }{RT}(\phi-\phi_{p})} \right)}  
\end{eqnarray}
where $ \Delta U =\sum_{i=1}^N(a_i-b_i) U_i+  \Delta zU_e$
and $ k_{eq} = e^{-\frac{\Delta U}{RT}}$ is used. 
If we assume 
\begin{equation}
\mathcal{R}_f = k_{f,0}\left(\frac{C_e}{c_{e,0}}\right)^{\Delta z}\Pi_{i=1}^N\left(\frac{C_i}{c_0}\right)^{a_i} e^{-\frac{\Delta Z F}{RT}\beta(\phi_{p}-\phi) },
\end{equation}
then the reaction rate function could be defined as follows
\begin{eqnarray}
\mathcal{R}& = &\mathcal{R}_f\left(1-e^{-\frac{\sum_{i=1}^N(a_i-b_i) \tilde{\mu}_i+\Delta z \tilde{\mu}_e}{RT}}\right)\nonumber\\
&=& k_{f,0}e^{-\frac{\Delta Z F}{RT}\beta(\phi_{p}-\phi) }\left(\frac{C_e}{c_{e,0}}\right)^{\Delta z}\Pi_{i=1}^N\left(\frac{C_i}{c_0}\right)^{a_i}  -k_{r,0}e^{\frac{\Delta Z F}{RT}(1-\beta)(\phi_{p}-\phi) } \Pi_{i=1}^N\left(\frac{C_i}{c_0}\right)^{b_i} \nonumber\\
&=& k_{f,0}e^{-\frac{\Delta Z F}{RT}\beta(\Delta\phi) }\left(\frac{C_e}{c_{e,0}}\right)^{\Delta z}\Pi_{i=1}^N\left(\frac{C_i}{c_0}\right)^{a_i}  -k_{r,0}e^{\frac{\Delta Z F}{RT}(1-\beta)(\Delta\phi) } \Pi_{i=1}^N\left(\frac{C_i}{c_0}\right)^{b_i} \nonumber\\
\end{eqnarray}
where we denote 
$\Delta\phi = \phi_{p}-\phi$ and $\beta$ is the so called the transfer coefficient commonly found in the Frumkin-Butler-Volmer Equation \cite{biesheuvel2009imposed,sekimoto2010stochastic,van2010diffuse,van2012frumkin,fraggedakis2021theory}.


 \textcolor{black}{It is important to realize that any analysis of boundary behavior requires detail appropriate for the mechanisms of flow across the boundary. In biological cases, these are often not yet known. In the physical case, details are significantly different in each of the many (quite diverse) systems that are described by the  Butler-Volmer equation of electrode reactions  \cite{biesheuvel2009imposed,sekimoto2010stochastic,van2010diffuse,van2012frumkin,fraggedakis2021theory}. The behavior of the systems is likely to be as diverse as the chemical reactions at the boundaries are themselves.}


In summary, we have the following system 
\begin{equation}\label{model_ele_sum}
\left\{
\begin{array}{ll}
\frac{\partial C_i}{\partial t} =-\nabla\cdot\boldsymbol{j}_i, & \mbox{in}~ \Omega \\[3mm]
-\nabla\cdot(\varepsilon_0\varepsilon_r \nabla \phi) = \sum_{i=1}^Nz_iFC_i, & \mbox{in}~ \Omega
\end{array}
\right.
\end{equation}
with boundary conditions 
\begin{equation}\label{bd_sum}
\left\{
\begin{array}{ll}
\boldsymbol{j}_i\cdot\boldsymbol{n} = -\gamma_i \mathcal{R}, i=1\cdots N,  & \mbox{on~} \Gamma,\\
\mathbb{C}_p\frac{d(\phi-\phi_p)}{dt} = -I_{ex} - \Delta zF\mathcal{R} = \frac{g}{F} (\tilde{\mu}_{ex}-\tilde{\mu}_e)- \Delta zF\mathcal{R} , &\mbox{on~} \Gamma\\
\mathcal{R} = k_{f,0}e^{-\frac{\Delta Z F}{RT}\beta(\Delta\phi) }\left(\frac{C_e}{c_{e,0}}\right)^{\Delta z}\Pi_{i=1}^N\left(\frac{C_i}{c_0}\right)^{a_i}  -k_{r,0}e^{\frac{\Delta Z F}{RT}(1-\beta)(\Delta\phi) } \Pi_{i=1}^N\left(\frac{C_i}{c_0}\right)^{b_i},&\mbox{on~} \Gamma\\
\boldsymbol{D}\cdot\boldsymbol{n}=  \mathbb{C}_p (\phi-\phi_p), &\mbox{on~} \Gamma\\
\boldsymbol{j}_i\cdot\boldsymbol{n} = 0,  \phi = 0,  &\mbox{on~}\partial\Omega/\Gamma.\\
\end{array}\right.
\end{equation}
where the  similar  boundary conditions are  used  in \cite{bazant2005current,biesheuvel2009imposed,yan2021adaptive}.

The reaction rate function could be written as 
\begin{equation}
\mathcal{R} = k_{f}\left(\frac{C_e}{c_{e,0}}\right)^{\Delta z}\Pi_{i=1}^N\left(\frac{C_i}{c_0}\right)^{a_i}  -k_{r}\Pi_{i=1}^N\left(\frac{C_i}{c_0}\right)^{b_i},
\end{equation}
where we denote
\begin{align}\label{reactionratefunction}
k_f =k_{f,0}e^{-\frac{\Delta Z F}{RT}\beta(\Delta\phi) },\\
k_r=k_{r,0}e^{\frac{\Delta Z F}{RT}(1-\beta)(\Delta\phi) },
\end{align} 
to include the effects of electric potential on the reaction rates. 

If we choose $g= \frac{I_{0}F}{ \tilde{\mu}_e -\tilde{\mu}_{ex}}$, then the input current is constant $I_0$ and the condition is changed to be $\mathbb{C}_p\frac{d(\phi-\phi_p)}{dt} = -I_{0} - \Delta zF\mathcal{R} $ which is the set of boundary condition found widely in the literature \cite{moya1995ionic,van2010diffuse,murphy1992numerical,bonnefont2001analysis}.  \textcolor{black}{This is based on Kirchhoff's Laws that the input current is either stored in the capacitor or consumed by the reaction. }



\begin{rmk}
Since $C_e^s = \frac{\mathbb{C}_p}{F}(\phi^s-\phi_p^s)$ is the charge density, it should be nonegative during evolution,which means $\phi^s>\phi_p^s\ge 0$ for $\forall t\ge 0$ . 
\end{rmk}
\begin{rmk}
If there is no chemical reaction on the surface $\Gamma$, the model degenerates to PNP with dynamic boundary condition
\begin{equation} 
\left\{
\begin{array}{ll}
	\frac{\partial C_i}{\partial t} =-\nabla\cdot\boldsymbol{j}_i, & \mbox{in}~ \Omega \\[3mm]
	-\nabla\cdot(\varepsilon_0\varepsilon_r \nabla \phi) = \sum_{i=1}^Nz_iFC_i, & \mbox{in}~ \Omega
\end{array}
\right.
\end{equation}
with boundary conditions 
\begin{equation} 
\left\{
\begin{array}{ll}
	\boldsymbol{j}_i\cdot\boldsymbol{n} = 0, ~~i=1\cdots N,  & \mbox{on~} \partial\Omega,\\
	\mathbb{C}_p\frac{d(\phi-\phi_p)}{dt} = -I_{ex}, &\mbox{on~} \Gamma,\\
	\boldsymbol{D}\cdot\boldsymbol{n}=  \mathbb{C}_p (\phi-\phi_p), &\mbox{on~} \Gamma,\\
	\phi = 0,  &\mbox{on~}\partial\Omega/\Gamma.\\
\end{array}\right.
\end{equation}

\end{rmk}

\subsection{A bi-reaction system} 

\begin{figure}
\centering
\includegraphics[width=0.6\textwidth]{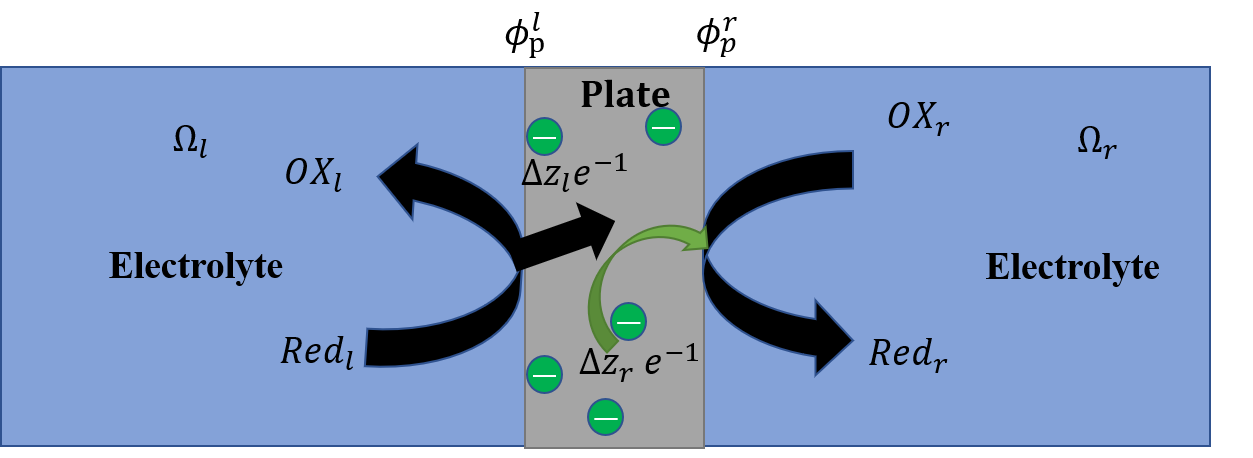}
\caption{Schematic of Bi-reaction system. On the right side of the plate, the reduction reaction consumes the electrons that are produced by the oxidation reaction on the left side of the plate.  }
\label{fig:bidomainsystem}
\end{figure}

 \textcolor{black}{We will now apply these general principles to a specific system of significant interest in biological applications, such as the electron transport chain (ETC) on the mitochondrial membrane \cite{guo2018structure,zhao2019mitochondrial,wikstrom2003water,wikstrom2018proton}.
The electron transport chain (ETC) is a series of biochemical reactions occurring in the inner mitochondrial membrane of eukaryotic cells or the plasma membrane of prokaryotic cells. It plays a pivotal role in cellular respiration, which is the process by which cells convert nutrients (such as glucose) into energy (in the form of ATP).
During the ETC, electrons transfer from electron donors (e.g., NADH and FADH2) to electron acceptors (e.g., oxygen) through a series of redox reactions. This electron transfer is accompanied by the pumping of protons (H+) across the inner mitochondrial or plasma membrane, creating an electrochemical gradient used to generate ATP. Oxidation occurs through a sequence of chemical reactions catalyzed by membrane proteins that connect electron flow on one side of a membrane to a chemical reaction on the other.
In essence, the reaction on one side of the plate generates electrons, which are subsequently consumed by the reaction on the other side. This scheme is visually represented in Fig. \ref{fig:bidomainsystem}, where the output of the left domain reaction is $|\Delta z_l|e^{-1}$ electrons and the input of the right domain is   $|\Delta z_r|e^{-1}$ electrons. }
 
 \textcolor{black}{To model this system, we set the domain $\Omega$ as a  union of two subdomains denoted as $\Omega_s$, where $s = l,r$, separated by the plate  $\Gamma$. The plate functions analogously to a cell membrane, allowing the conduction of electrons along with a displacement current through the effective membrane capacitance $\mathbb{C}_p$.
The thickness of the plate is negligible when compared to the size of the domain.
Within these two compartments $\Omega_s$, where $s=l,r$, there exist a total of $N$ species, represented as $\{X_1^{z_1}, X_2^{z_2},\cdots, X_N^{z_N}\}$, with associated reactions as follows:} 
 \textcolor{black}{\begin{equation} 
a^s_1 X_1^{z_1}+a^s_2 X_2^{z_2}+\cdots+a^s_N X_N^{z_N}+\Delta z^s e^{-1}\overset{k^s_{f,0}}{\underset{k^s_{r,0}}{\rightleftharpoons}}  b^s_1 X_1^{z_1}+b^s_2 X_2^{z_2}+\cdots+b^s_N X_N^{z_N},
\end{equation}}
 \textcolor{black}{where $k^s_{f,0}$ and $k^s_{r,0}$ are the intrinsic forward and backward reaction rates.  
In the left region $\Omega_l$, where $\Delta z^l<0$, this signifies the forward reaction as oxidation, a process that generates electrons. In the right compartment $\Omega_r$, where $\Delta z^r>0$, this indicates that the forward reaction is a reduction process, consuming electrons.}   

In the electrolyte region $\Omega_s$, we have the following kinematic assumption 
\begin{equation} 
\left\{
\begin{array}{l}
\frac{\partial C^s_i}{\partial t} =-\nabla\cdot \bj_i^s,\\[3mm]
-\nabla\cdot(\varepsilon_0\varepsilon_r \nabla \phi^s) =   \sum_{i=1}^N z_i FC_i^s
\end{array}
\right.
\end{equation}
with the boundary  conditions 
 
\begin{equation} 
\left\{
\begin{array}{ll}
\boldsymbol{j}^s_i\cdot\boldsymbol{n}_s =  -\gamma_i \mathcal{R}^s,  & \mbox{on~} \Gamma,\\
\mathbb{C}_p\frac{d(\phi^l-\phi^l_p)}{dt} = -\Delta z^lF\mathcal{R}_l+I_{ex}, \mathbb{C}_p\frac{\phi^r-\phi^r_p}{dt} = -\Delta z^r\mathcal{R}_rF-I_{ex},  & \mbox{on~} \Gamma,\\
\boldsymbol{j}^s_i\cdot\boldsymbol{n} = 0, ~  &\mbox{on~}\partial\Omega\setminus\Gamma,
\end{array}\right.
\end{equation} 
 \textcolor{black}{where the second equation is based on Kirchhoff's Laws and $I_{ex}$ represents the current passing through the plate due to electron transportation.}

The total energy is defined as  \textcolor{black}{ the sum of energy  within the two subdomains and on the interface  }
\begin{eqnarray}
E &=& E_{l}+E_{r} +E_{\Gamma}\nonumber\\
&= &\sum_{i=1}^N \int_{\Omega^l}\left\{ RTC_i^l\left(\ln{\left(\frac{C_i^l}{c_0}\right)}-1 \right) +C_i^lU_i\right\}dx 
+
\int_{\Omega^l} \frac{\boldsymbol{D}^l\cdot\boldsymbol{E}^l}{2}dx\nonumber\\
&&+\sum_{i=1}^N \int_{\Omega^r}\left\{ RTC_i^r\left(\ln{\left(\frac{C_i^r}{c_0}\right)}-1 \right) +C_i^rU_i\right\}dx 
+
\int_{\Omega^r} \frac{\boldsymbol{D}^r\cdot\boldsymbol{E}^r}{2}dx\nonumber\\
&&
+\int_{\Gamma} \left\{ RTC_e^l\left(\ln{\left(\frac{C_e^l}{c_{e,0}}\right)}-1 \right) +C_e^lU_e^l\right\}dx +\frac 1 2 \int_{\Gamma} \mathbb{C}_p(\phi^l-\phi_p^l)^2 dS\nonumber\\
&&
+\int_{\Gamma} \left\{ RTC_e^r\left(\ln{\left(\frac{C^r_e}{c_{e,0}}\right)}-1 \right) +C_e^rU_e^r\right\}dx +\frac 1 2 \int_{\Gamma} \mathbb{C}_p(\phi^r-\phi_p^r)^2 dS.
\end{eqnarray}
 \textcolor{black}{Similarly, } the dissipative function is defined as 
\begin{eqnarray}
\Delta&=& \Delta_l+\Delta_r +\Delta_{\Gamma}\nonumber\\
&=&\int_{\Omega^l} \sum_{j=1}^N\frac{D_iC_i^l}{RT}|\nabla\tilde\mu_i^l|^2dx+ \int_{\Gamma} RT\mathcal{R}^l\ln \left(\frac{\mathcal{R}_f^l}{\mathcal{R}_r^l}\right)  dx\nonumber\\
&&+\int_{\Omega^r} \sum_{j=1}^N\frac{D_iC_i^r}{RT}|\nabla\tilde\mu_i^r|^2dx+ \int_{\Gamma} RT\mathcal{R}^r\ln \left(\frac{\mathcal{R}_f^r}{\mathcal{R}_r^r}\right)  dx+\int_{\Gamma}\frac{g(C_e^l)}{F^2}(\tilde{\mu}_e^l-\tilde{\mu}_e^r)^2dS.
\end{eqnarray}
For the boundary input energy power, it is assumed  to be 
\begin{equation}
\mathcal{P}_{E,\partial \Omega} = -\int_{\partial\Omega}\phi_{ref}\frac{\partial \mathbf{D}\cdot\mathbf{n}}{\partial t} dx. 
\end{equation}

Taking the time derivative of the total energy yields 
 \textcolor{black}{\begin{eqnarray}\label{eqn:dedtoneside}
 \frac{dE}{dt}-\mathcal{P}_{E,\partial\Omega} 
&=&\int_{\Omega_l}\sum_{i=1}^N \nabla\tilde{\mu}_i^l\cdot \boldsymbol{j}_i^l dx -\int_{\Gamma}(-\sum_{i=1}^N \tilde{\mu}_i^l \gamma_i\mathcal{R}^l+\tilde{\mu}_e^l\Delta z^l)\mathcal{R}_l dS -\int_{\Gamma}\phi_l\frac{\partial }{\partial t}(\boldsymbol{D}^l\cdot\boldsymbol{n}^l-FC_e^l)dS \nonumber \\
&& +\int_{\Omega_r}\sum_{i=1}^N \nabla\tilde{\mu}_i^r\cdot \boldsymbol{j}_i^r dx -\int_{\Gamma}\left(-\sum_{i=1}^N \tilde{\mu}_i^r \gamma_i+\Delta z^r\tilde{\mu}_e^r\right)\mathcal{R}^r dS\nonumber\\
&&  -\int_{\Gamma}\phi_r\frac{\partial }{\partial t}(\boldsymbol{D}^r\cdot\boldsymbol{n}^r-FC_e^r)dS + \int_{\Gamma}(\tilde{\mu}_e^r
-\tilde{\mu}_e^l)J_e-\int_{\partial\Omega}(\phi-\phi_{ref})\frac{\partial \boldsymbol{D}\cdot\boldsymbol{n}}{\partial t}dS\nonumber\\
&=&  -\Delta.
\end{eqnarray}}

Matching terms yield 
\begin{equation}
\left\{
\begin{array}{ll}
\boldsymbol{j}^s_i = -\frac{D^s_i C^s_i}{RT}\nabla\tilde{\mu}^s_i, & \mbox{in~}\Omega^s\\
\mathcal{R}^l = k^l_{f,0}e^{-\frac{\Delta Z F}{RT}\beta(\Delta\phi^l) }\Pi_{i=1}^N\left(\frac{C^l_i}{c_0}\right)^{a_i}  -k^l_{r,0}e^{\frac{\Delta z^l F}{RT}(1-\beta)(\Delta\phi^l) }\Pi_{i=1}^N\left(\frac{C^l_e}{c_{e,0}}\right)^{-\Delta z^l}\left(\frac{C^l_i}{c_0}\right)^{b_i} & \mbox{on~}\Gamma^l\\
\mathcal{R}^r = k^r_{f,0}e^{-\frac{\Delta Z F}{RT}\beta(\Delta\phi^r) }\left(\frac{C^r_e}{c_{e,0}}\right)^{\Delta z^r}\Pi_{i=1}^N\left(\frac{C^r_i}{c_0}\right)^{a_i}  -k^r_{r,0}e^{\frac{\Delta z^r F}{RT}(1-\beta)(\Delta\phi^r) }\Pi_{i=1}^N\left(\frac{C^r_i}{c_0}\right)^{b_i} & \mbox{on~}\Gamma^r\\
\boldsymbol{D}^s\cdot\boldsymbol{n}^s = C_e^sF  = \mathbb{C}_p(\phi^s-\phi_p^s), &\mbox{on} ~ \Gamma\\
J_e = -\frac{g}{F^2} (\tilde{\mu}_e^r-\tilde{\mu}_e^l), &\mbox{on} ~ \Gamma\nonumber\\
\phi = \phi_{ref}, &\mbox{on}~\partial\Omega/\Gamma.
\end{array}
\right.
\end{equation}

In summary,  the bidomain reaction system is 
\begin{equation} 
\left\{
\begin{array}{ll}
\frac{\partial C^s_i}{\partial t} = \nabla\cdot ( \frac{D^s_i C^s_i}{RT}\nabla\tilde{\mu}^s_i), &\mbox{in~}\Omega^s\\[3mm]
-\nabla\cdot(\varepsilon_0\varepsilon_r \nabla \phi^s) =   \sum_{i=1}^N z_i FC_i^s, &\mbox{in~}\Omega^s
\end{array}
\right.
\end{equation}
for the boundary  conditions 
\begin{equation} 
\left\{
\begin{array}{ll}
\boldsymbol{j}^s_i\cdot\boldsymbol{n}_s =  -\gamma_i \mathcal{R}^s,  & \mbox{on~} \Gamma,\\
\mathcal{R}^l = k^l_{f,0}e^{-\frac{\Delta Z F}{RT}\beta(\Delta\phi^l) }\Pi_{i=1}^N\left(\frac{C^l_i}{c_0}\right)^{a_i}  -k^l_{r,0}e^{\frac{\Delta z^l F}{RT}(1-\beta)(\Delta\phi^l) }\Pi_{i=1}^N\left(\frac{C^l_e}{c_{e,0}}\right)^{-\Delta z^l}\left(\frac{C^l_i}{c_0}\right)^{b_i} & \mbox{on~}\Gamma^l\\
\mathcal{R}^r = k^r_{f,0}e^{-\frac{\Delta Z F}{RT}\beta(\Delta\phi^r) }\left(\frac{C^r_e}{c_{e,0}}\right)^{\Delta z^r}\Pi_{i=1}^N\left(\frac{C^r_i}{c_0}\right)^{a_i}  -k^r_{r,0}e^{\frac{\Delta z^r F}{RT}(1-\beta)(\Delta\phi^r) }\Pi_{i=1}^N\left(\frac{C^r_i}{c_0}\right)^{b_i} & \mbox{on~}\Gamma^r\\
\mathbb{C}_p\frac{d(\phi^l-\phi^l_p)}{dt} = -\Delta z^lF\mathcal{R}^l +\frac {g } F(\tilde{\mu}_e^r-\tilde{\mu}_e^l), \mathbb{C}_p\frac{d(\phi^r-\phi^r_p)}{dt} = -\Delta z^rF\mathcal{R}^r -\frac {g } F (\tilde{\mu}_e^r-\tilde{\mu}_e^l),  & \mbox{on~} \Gamma,\\
\boldsymbol{D}^s\cdot\boldsymbol{n}^s   = \mathbb{C}_p(\phi^s-\phi_p^s),  & \mbox{on~} \Gamma,\\
\boldsymbol{j}^s_i\cdot\boldsymbol{n} = 0, 
\phi=\phi_{ref}~  &\mbox{on~}\partial\Omega\setminus\Gamma.
\end{array}\right.
\end{equation} 



 \textcolor{black}{ 
One of the crucial properties of cell membranes is their voltage-gated mechanisms \cite{catterall2000ionic}. These mechanisms encompass a category of membrane proteins, known as ion channels, which are highly responsive to variations in the electric potential difference across a cell's membrane. These ion channels play a pivotal role in various physiological processes within neurons, muscles, and other excitable cells. Their significance lies in their ability to initiate and propagate electrical signals, including action potentials, which are fundamental for nerve cell communication \cite{bezanilla2006action} and muscle contraction \cite{nieves2018regulation}. Various models, such as the Hodgkin-Huxley model \cite{hodgkin1952quantitative}, have been proposed to describe voltage-gated mechanisms.}

 \textcolor{black}{For the sake of simplicity, we model these mechanisms by introducing a potential difference threshold, denoted as $\theta_0$, using a hard switch }
\begin{equation}\label{hardswitch}
g = \left\{\begin{array}{ll}
g_0  & \mbox{if~} \phi_p^r-\phi_p^l >\theta_0;\\
0   & \mbox{otherwise},   
\end{array}
\right.
\end{equation}
or a soft one
\begin{equation}\label{softswitch}
g = g_0 \tanh \left(\frac{(\phi_p^r-\phi_p^l)-\theta_0}{\epsilon}\right),
\end{equation}
 \textcolor{black}{where $g_0$ is the maximum conductance  and $\epsilon$ is the relaxation. }
Then,  when the difference of potential between the two sides of the plate is large enough, the plate becomes conductive and the electron is transported.

\section{Results} \label{section:simulation}

In the simulation, we consider the 1D  computation domain and the plate is in the middle. On the left side of the  plate, we have
 \textcolor{black}{$$X_1^{+}+X_2^{2-} \overset{k^l_f}{\underset{k^l_r}{\rightleftharpoons}}  X_3+e^{-}.$$
On the  right side of  the plate,
$$X_4^{2+}+X_5^{-} +e^{-} \overset{k^r_f}{\underset{k^r_r}{\rightleftharpoons}} X_6.$$ }
 \textcolor{black}{Let's introduce the following characteristic values: $L = 4.3\times10^{-1}~ \mathrm{\mu m}$, $c_0=1~ \mathrm{mM}$, $c_{e,0} = \frac{\mathbb{C}_p RT}{F^2} = 4.35\times 10^2 ~ \mathrm{mol/\mu m^2}$, $D^* = 10^{-9}~  \mathrm{m^2/s}$, $t^*=\frac{L^2}{D^*} = 1.85\times 10^{-4}~  \mathrm{s}$, $\frac{RT}{F}=25.6~  \mathrm{mv}$, and $k_f^* = 2.33\times 10^4 ~ \mathrm{M/s}$. These values represent the characteristic length, concentration, electron density, diffusion constant, time, electric potential, and reaction rate, respectively. With these characteristic values, we can derive the dimensionless system as follows. }
\begin{equation} \label{bidomain system}
\left\{
\begin{array}{l}
\frac{\partial C^s_i}{\partial t} = \nabla\cdot (D_i^s (\nabla C_i^s+z_i C_i^s\nabla\phi^s)),\\[3mm]
-\nabla\cdot(\delta^2\nabla \phi^s) =   \sum_{i=1}^N z_i C_i^s,
\end{array}
\right.
\end{equation}
with the boundary conditions  
\begin{equation} \label{bidomain system bd}
\left\{
\begin{array}{ll}
\boldsymbol{j}^s_i\cdot\boldsymbol{n}_s =  -\gamma_i \zeta\mathcal{R}^s,  & \mbox{on~}  \Gamma,\\
 -\nabla\phi_s\cdot\boldsymbol{n}_s = \lambda_s (\phi^s-\phi_p^s),  & \mbox{on~} \Gamma,\\    
~\frac{\partial}{\partial t}(-\nabla\phi_l\cdot\boldsymbol{n_l}) +\frac{\zeta  }{\delta^2}\Delta z_l \mathcal{R}_l  = g \left(\ln(\frac{C_e^r}{C_e^l})+\phi_p^l-\phi_p^r\right) ,   & \mbox{on~} \Gamma,\\    
\frac{\partial}{\partial t}(-\nabla\phi_r\cdot\boldsymbol{n_r}) +\frac{\zeta  }{\delta^2}\Delta z_r \mathcal{R}_r =-g \left(\ln(\frac{C_e^r}{C_e^l})+\phi_p^l-\phi_p^r\right), & \mbox{on~} \Gamma,\\
\boldsymbol{j}^s\cdot\boldsymbol{n} = 0, \phi^s = 0, &\mbox{on~}\partial\Omega/\Gamma,
\end{array}\right.
\end{equation} 
where  
  \textcolor{black}{ \begin{eqnarray}
  \mathcal{R}^l &=&  k_{f,0}^l e^{-\Delta Z_l \beta(\phi_p^l-\phi^l) }\Pi_{i=1}^N\left(C^l_i\right)^{a_i} -k_{r,0}^le^{ \Delta Z_l(1-\beta)(\phi_p^l-\phi^l) }(C_e^l)^{-\Delta z^l}\Pi_{i=1}^N\left( C^l_i \right)^{b_i} \nonumber\\
    &=& k_{f}^l\Pi_{i=1}^N\left(C^l_i\right)^{a_i} -k_{r}^l(C_e^l)^{-\Delta z^l}\Pi_{i=1}^N\left( C^l_i \right)^{b_i}\nonumber\\
     \mathcal{R}^r &=& k_{f,0}^re^{-\Delta Z_r \beta(\phi_p^r-\phi^r) } (C_e^r)^{\Delta z^r}\Pi_{i=1}^N\left(C^r_i\right)^{a_i} -k_{r,0}^re^{ \Delta Z_r(1-\beta)(\phi_p^r-\phi^r) }\Pi_{i=1}^N\left( C^r_i \right)^{b_i} .\nonumber\\
    &=&  k_{f}^r (C_e^r)^{\Delta z^r}\Pi_{i=1}^N\left(C^r_i\right)^{a_i} -k_{r}^r\Pi_{i=1}^N\left( C^r_i \right)^{b_i}\nonumber
\end{eqnarray}}
  \textcolor{black}{with $k_f^s =  k_{f,0}^s e^{-\Delta Z_s \beta(\phi_p^s-\phi^s) }$  and $k_r^s = k_{r,0}^se^{ \Delta Z_s(1-\beta)(\phi_p^s-\phi^s) }.$}
 
Here  $\delta = \frac{\lambda_D}{L}$ is the ratio between Debye length $\lambda_D =\sqrt{\frac{RT\epsilon_0\epsilon_r}{F^2c_0}}$ and macroscale length;  
$\zeta = \frac{k_f^*L}{D^*c_0}$ is the ratio between the reaction time and diffusion time; $\lambda_s =\frac{C_sL}{\epsilon_0\epsilon_r}$ is the ratio between  macroscale length and effective width for the Stern layer;
and $g=\frac{gt^*L}{\epsilon_0\epsilon_r}$ is the ratio between the diffusion time and electric conduction time.

The initial concentrations and dimensionless parameters  are set to be $C_1(t=0) = 1$, $C_2(t=0)=\frac 1 2$, $C_3(t=0)=0$, $C_4 (t=0)= \frac 1 2$, $C_5(t=0)=1$, $C_6(t=0)=0$, $\delta = 0.1$, $ g=0.1$, $k_{r,0}^l=k_{r,0}^r=0.001$.  
 \textcolor{black}{
The results in the remaining part of the paper are presented in a dimensionless format. }

\subsection{ \textcolor{black}{Equal  intrinsic reaction rates}}
\begin{figure}[h]
\centering
\begin{subfigure}[]{
		\includegraphics[width=0.35\textwidth]{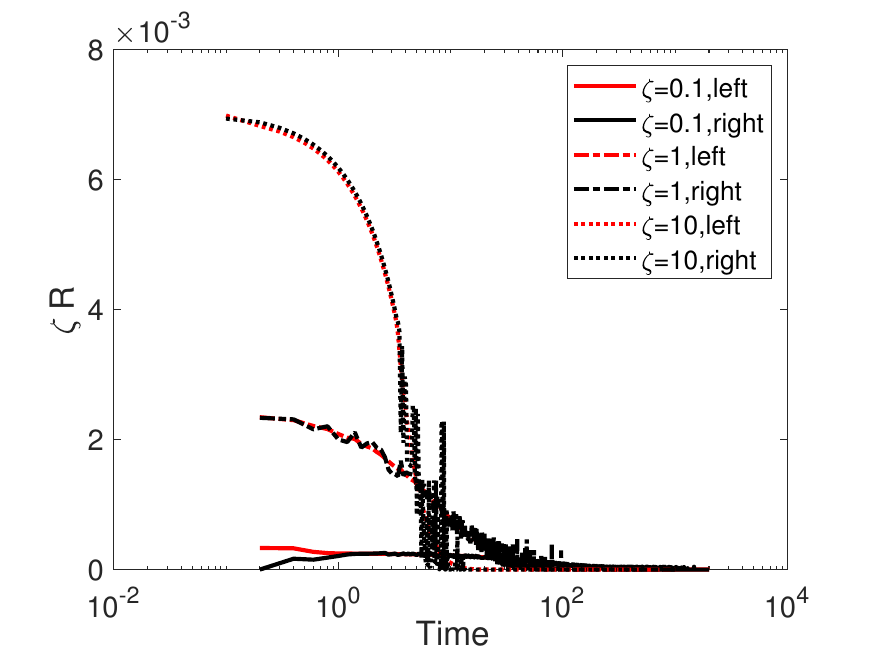}}
\end{subfigure}
\begin{subfigure}[]{
		\includegraphics[width=0.35\textwidth]{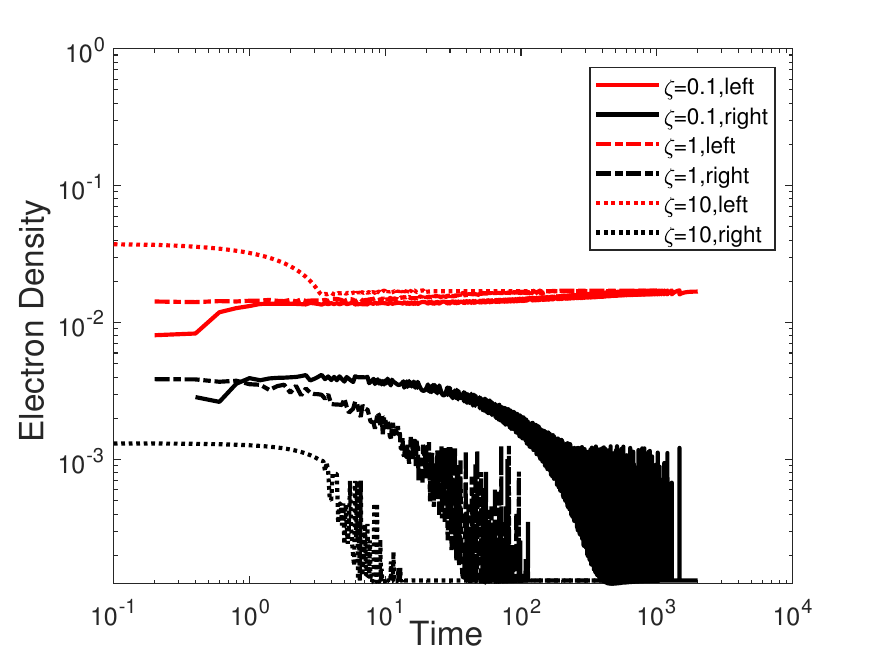}}
\end{subfigure}
\begin{subfigure}[]{
		\includegraphics[width=0.45\textwidth]{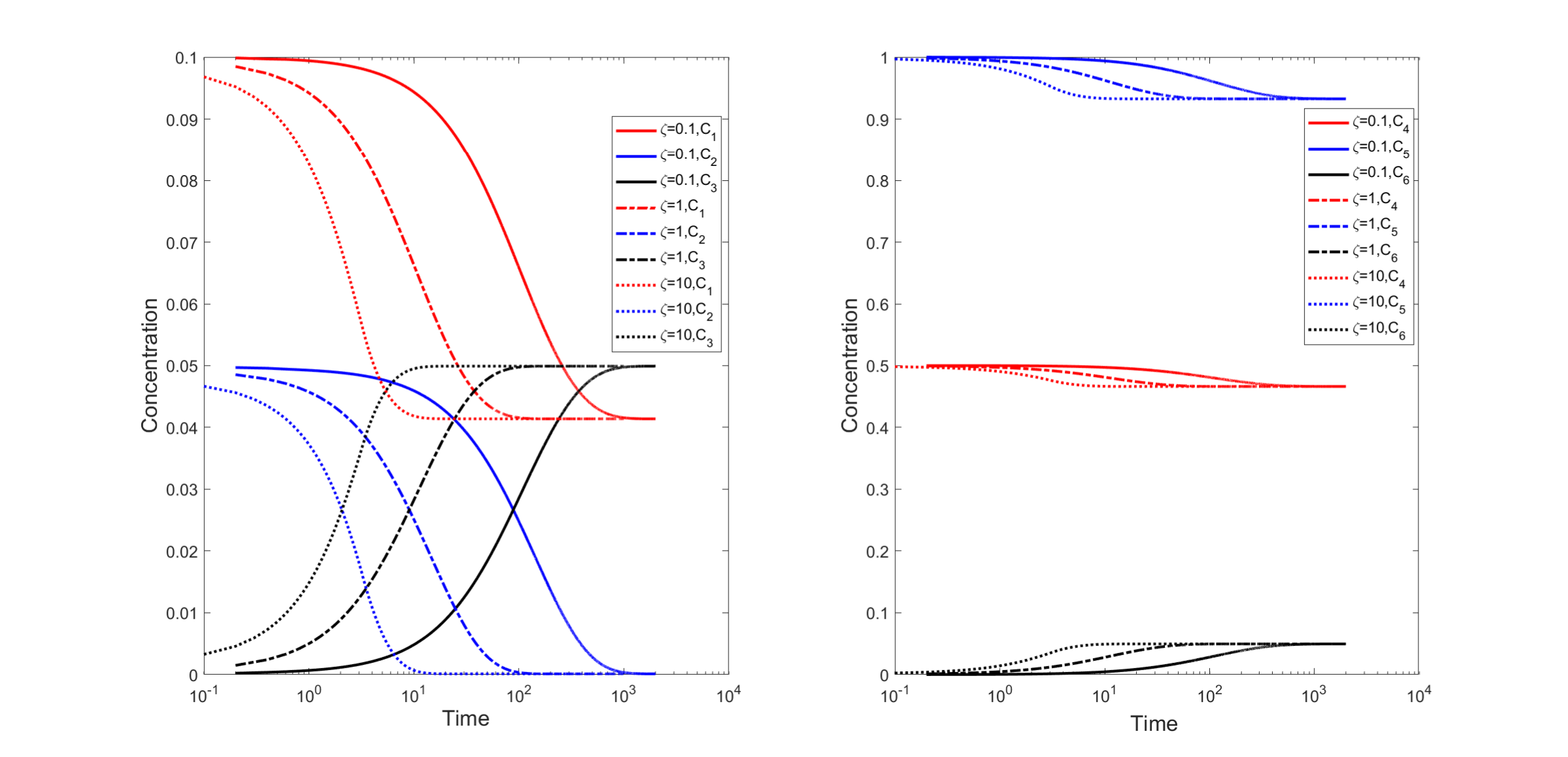}}
\end{subfigure}
\begin{subfigure}[]{
		\includegraphics[width=0.45\textwidth]{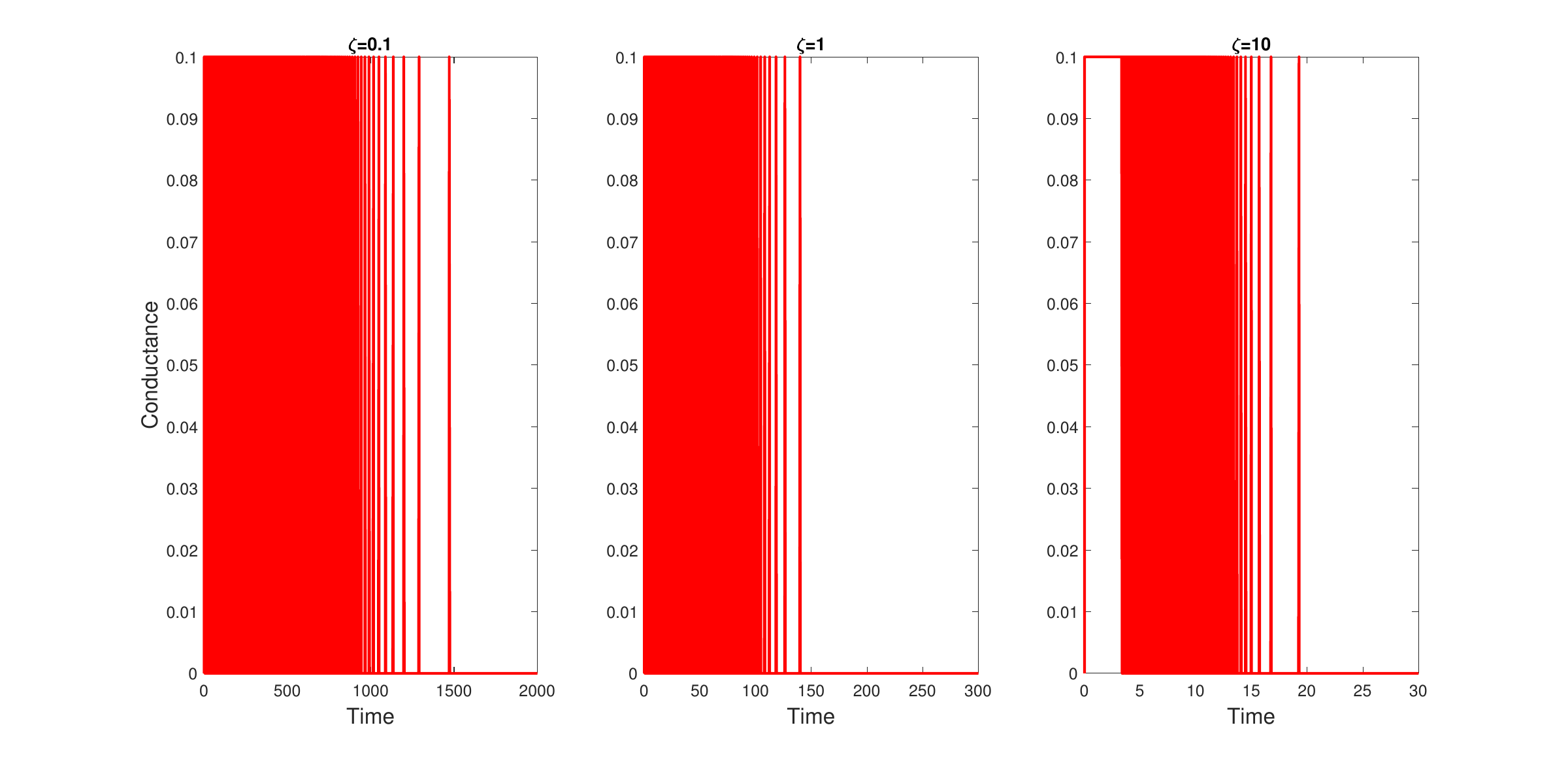}}
\end{subfigure}	

\caption{  \textcolor{black}{Dynamics evolution of  (a) Reaction rates (b) Electron Density (c) Substance concentrations and (d) Conductance changes over time with the same intrinsic reaction rate $k_{f,0}$ on both sides of the plate and different timescale ratio $\zeta$.  The results show that the reaction on the left side of the plate is responsible for generating
electrons, which are subsequently consumed by the reaction on the right side. The larger $\zeta$ is, the earlier the switch turns off and the reaction achieves equilibrium. Dimensional units are used as defined in the beginning of the Results Section.} }
\label{fig:lreqcm001time}
\end{figure}

We first consider  \textcolor{black}{the simplified case} that the left and right reactions with the same  \textcolor{black}{intrinsic} reaction rate, $k_{f,0}^l=k_{f,0}^r = 1$. 
Fig. \ref{fig:lreqcm001time} shows the results of the dynamics of reaction rates, electron density, concentrations of substances, and conductance over time with different  \textcolor{black}{time-scale ratio}   $\zeta=0.1$ (diffusion dominates),  $\zeta=1$ (diffusion-reaction equal), $\zeta=10$ (reaction dominates). 

 \textcolor{black}{As shown in Fig. \ref{fig:lreqcm001time}a,  the left reaction can be seen as a source---an input--- that continually produces electrons until one of the reactants, in this case, $X_2$, is depleted.  $X_2$ is the output. (\ref{fig:lreqcm001time}c). The electrons generated on the left are transported to the right side by the electrochemical potential difference, subsequently driving the reaction on the right side, the output side. Eventually, the output reaction achieves equilibrium after the switch turns off. ( Fig. \ref{fig:lreqcm001time}b).}

As time-scale ratio $\zeta$ increases,   the reaction rate also increases (Fig. \ref{fig:lreqcm001time}a).  The reactants on the left plates are consumed much faster  (Fig. \ref{fig:lreqcm001time}c left) and more electrons accumulate on the left side of the plates  (Fig. \ref{fig:lreqcm001time}b).  The electric potential difference between two sides of the plates $\phi_p^r-\phi_p^l$ increases and opens the gate so that electrons are transported to the right side of the plate (Fig. \ref{fig:lreqcm001time}b,d). Then the right reaction starts \ref{fig:lreqcm001time}c right). Due to the transport of electrons, the difference in potential between two sides of the plate decreases. The switch then turns off producing the gating dynamics shown in Fig. \ref{fig:lreqcm001time}d. Since the total amounts of reactants on the left region are limited, the left reaction stops after a   period of time because it runs out of electrons. Without a supply of electrons, the switch turns and the right-side reaction stops. The larger $\zeta$ is, the earlier the reaction stops and the switch turns off.  \textcolor{black}{The molecular mechanism of the switch remains to be investigated. It resembles the gate of ion channels and may have the same molecular basis \cite{gadsby2009ion}.}

\begin{figure}[!h]
\centering
\includegraphics[width=0.7\textwidth]{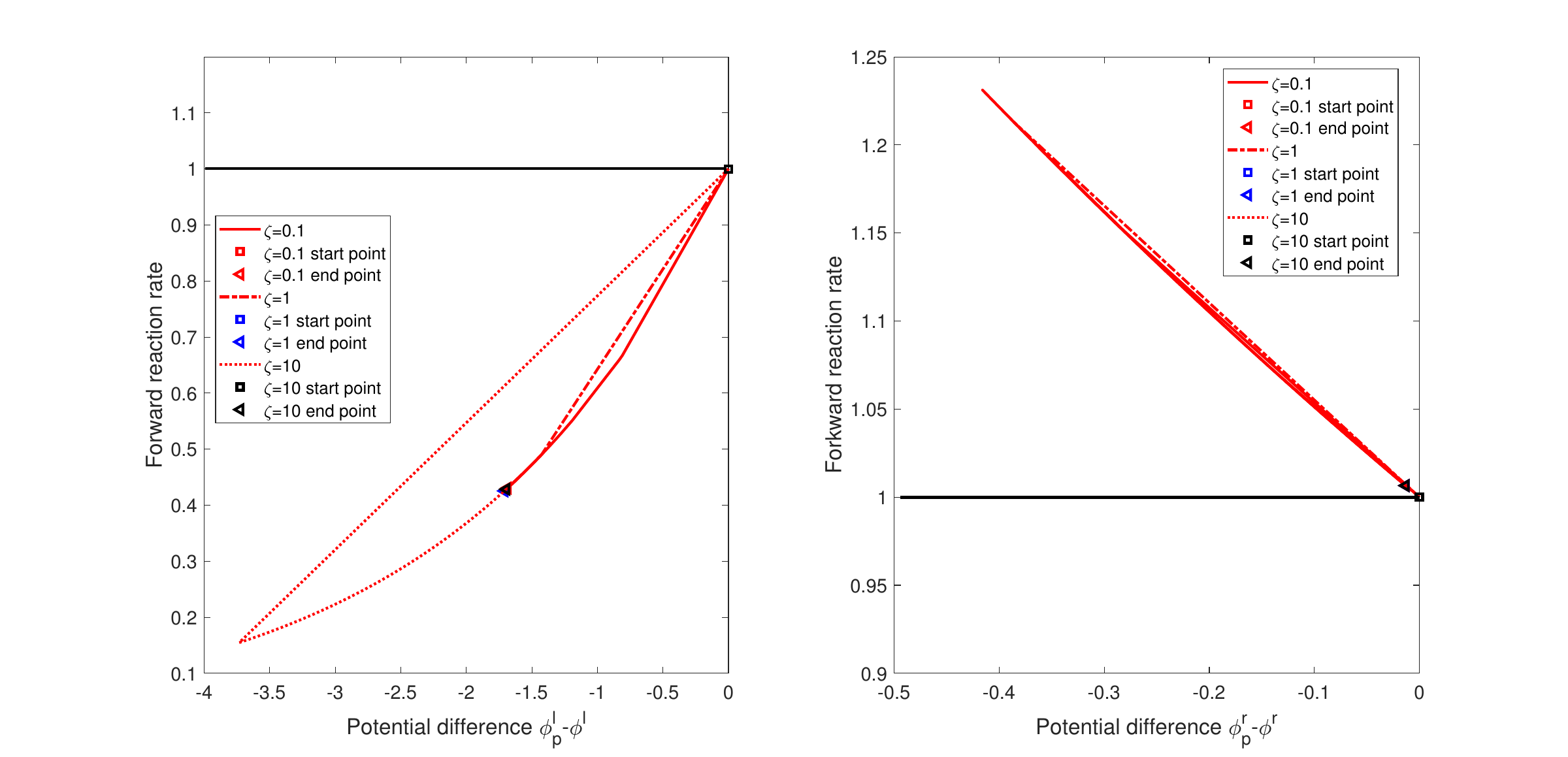}
\caption{ \textcolor{black}{Phase portraits} of  Reaction rate $k_f^s$ and electric potential difference $\phi_p^l-\phi_l$. Left: left plate; Right: right  plate. The black line is the  \textcolor{black}{intrinsic} forward reaction rate value $k^{s}_{f,0}=1, s= l,r$. The squares are the start points and the triangles are the endpoints of the  \textcolor{black}{phase portraits. It shows that the electric potential inhibits the oxidation reaction while accelerating the reduction reaction. Dimensional units are used as defined in the beginning of the Results Section.}}
\label{fig:lreqcm001ratevspotential}
\end{figure}

In Fig. \ref{fig:lreqcm001ratevspotential}, we show   \textcolor{black}{how the reaction rate $k_f^s$ changes with electric potential difference $\Delta\phi^s = \phi_p^s-\phi_s$, i.e. the phase portraits, in Eq.\eqref{reactionratefunction}.} 
The squares and triangles are the start points and end points of the  \textcolor{black}{phase portraits}, respectively. $k_{f,0}^s$ is the  \textcolor{black}{intrinsic rate constant} shown as the solid black line in the figure. When the electric potentials  \textcolor{black}{vary} with time, the forward reaction rates on the plate  \textcolor{black}{vary} correspondingly. Since $\Delta z_l<0$ and $\phi_p^l<\phi_s^l$, then the reaction rate on the left plate is smaller than the reference value $k_{f,0}^l$. On the right plate, since $\Delta z_r>0$, the reaction rate is always greater than the reference value $k_{f,0}^r$.   \textcolor{black}{ In summary, the electric potential inhibits the oxidation reaction on the left-hand input side, while accelerating the reduction reaction on the output right-hand side. }  

\begin{figure}[h]
\centering
\begin{subfigure}[]{
		\includegraphics[width=0.36\textwidth]{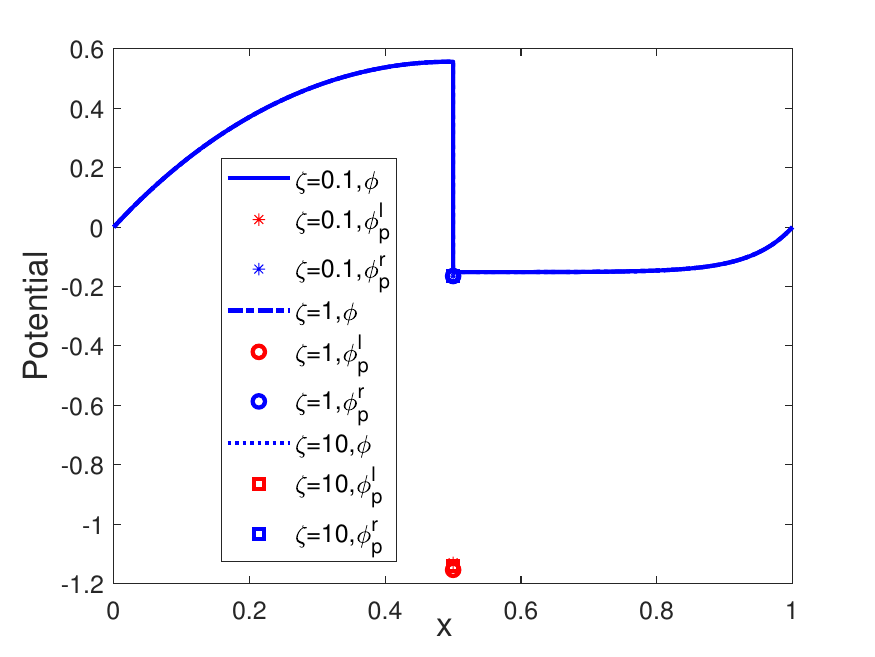}}
\end{subfigure}
\begin{subfigure}[]{
		\includegraphics[width=0.55\textwidth]{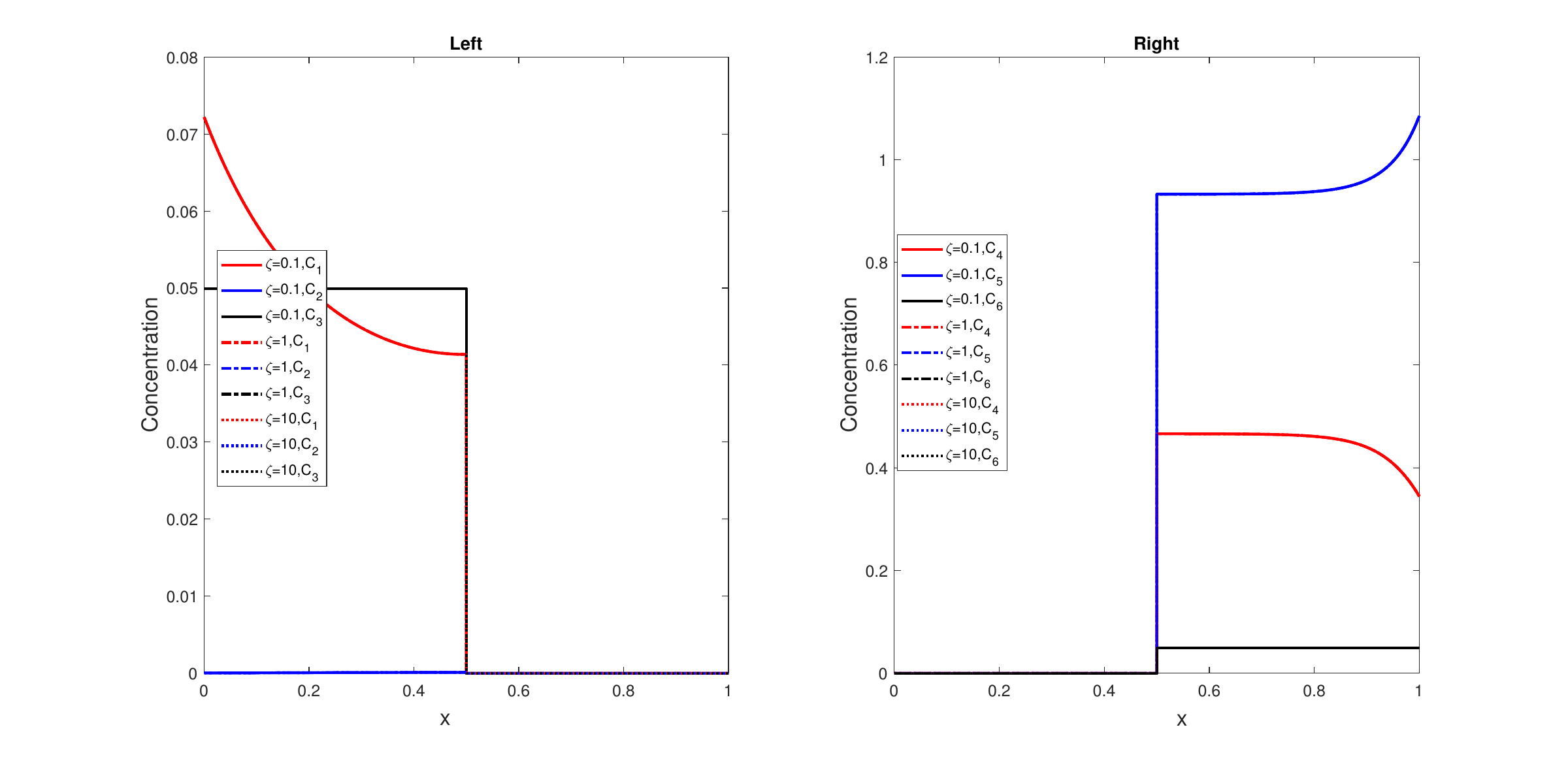}}
\end{subfigure}
\caption{   \textcolor{black}{Space distributions of    (a) Electric potential; (b) Substance concentrations at equilibrium with the same intrinsic reaction rate $k_{f,0}$ on both sides of the plate and different timescale ratio $\zeta$. Dimensional units are used as defined in the beginning of the Results Section.}   }
\label{fig:lreqkappa001space}
\end{figure}

The spatial distributions of electric potential and  \textcolor{black}{concentrations} in two regions at equilibrium are shown in Fig. \ref{fig:lreqkappa001space}.  For different $\zeta$, the equilibrium is almost the same. The electric potentials on two sides of the plate $\phi_p^s$ are lower than the potentials of the electrolyte $\phi^s$ due to the existence of capacitors. The left plate potential is lower than the potential on the right plate due to the accumulation of electrons.  \textcolor{black}{Since  $C_3$ and $C_6$ are neutral particles, they are uniformly distributed  \textcolor{black}{in sapce}. The distributions of cations and anions are changed by the electric field. }

\begin{figure}[ht]
\centering
\begin{subfigure}[]{
		\includegraphics[width=0.25\textwidth]{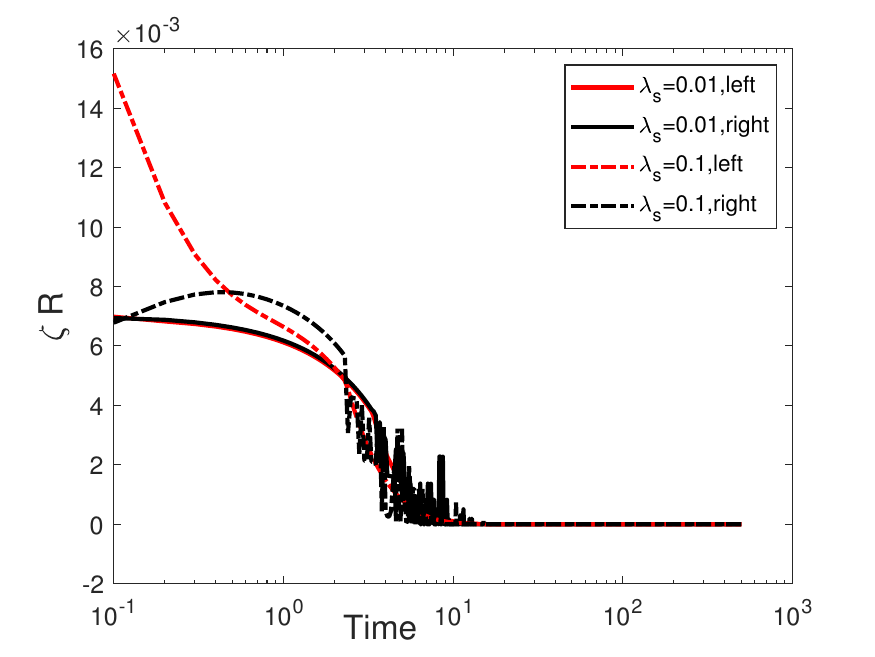}}
\end{subfigure}
\begin{subfigure}[]{
		\includegraphics[width=0.25\textwidth]{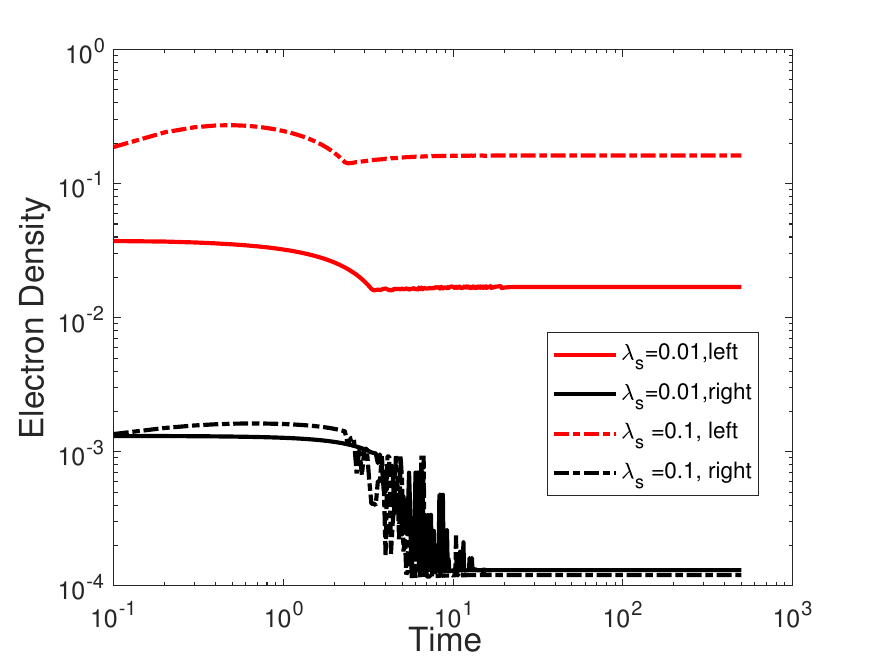}}
\end{subfigure}
\begin{subfigure}[]{
		\includegraphics[width=0.55\textwidth]{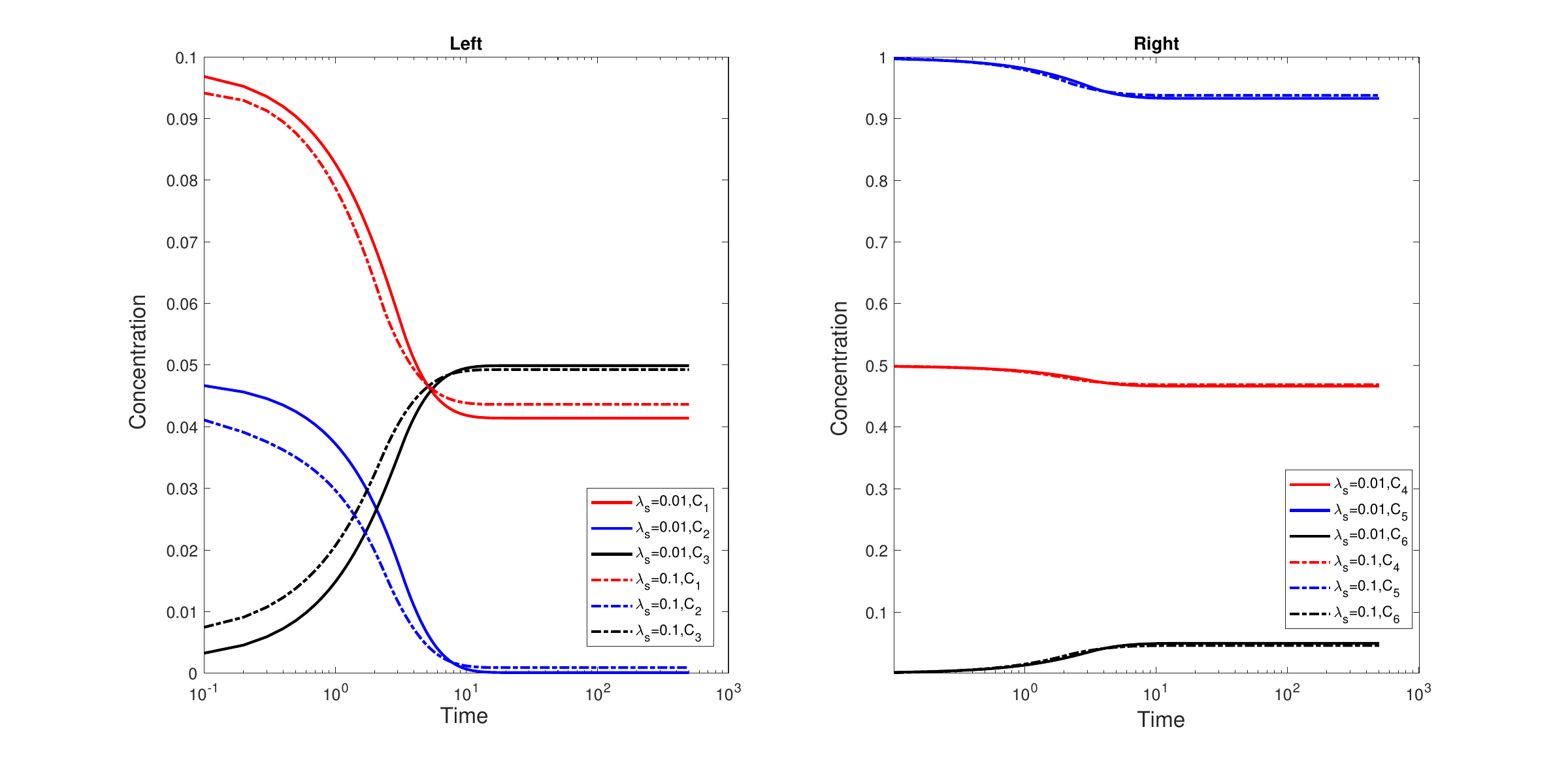}}
\end{subfigure}	
\caption{  \textcolor{black}{Dynamics evolution of  (a) Reaction rates (b) Electron Density (c) Substance concentrations over time with the same intrinsic reaction rate $k_{f,0}$ on both sides of the plate and different capacitance $\lambda_s$. Here $\zeta = 10$ is fixed. Dimensional units are used as defined in the beginning of the Results Section.}  }
\label{fig:lreqdiffcmtime}
\end{figure}

\begin{figure}[h]
\centering
\begin{subfigure}[]{
		\includegraphics[width=0.36\textwidth]{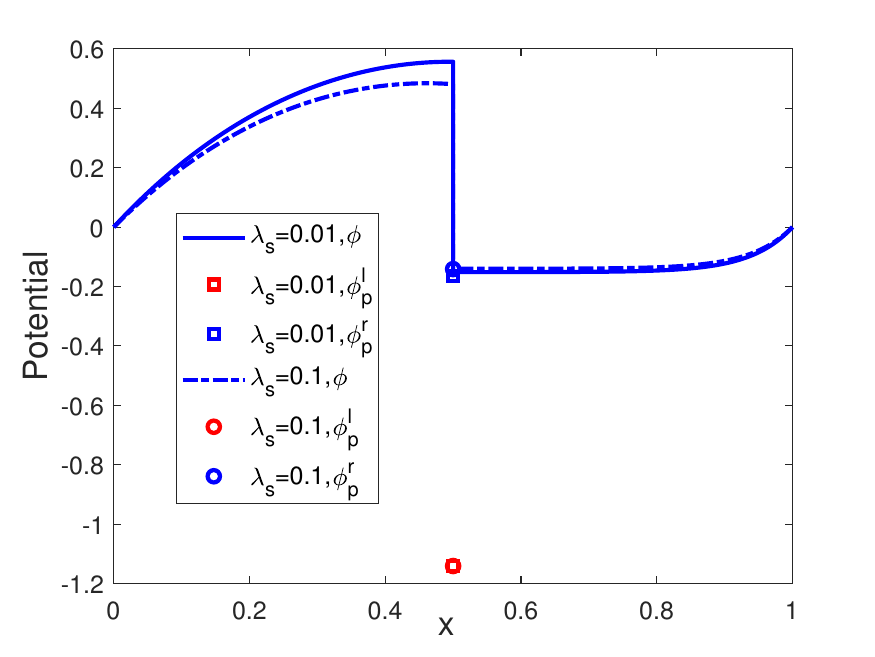}}
\end{subfigure}
\begin{subfigure}[]{
		\includegraphics[width=0.55\textwidth]{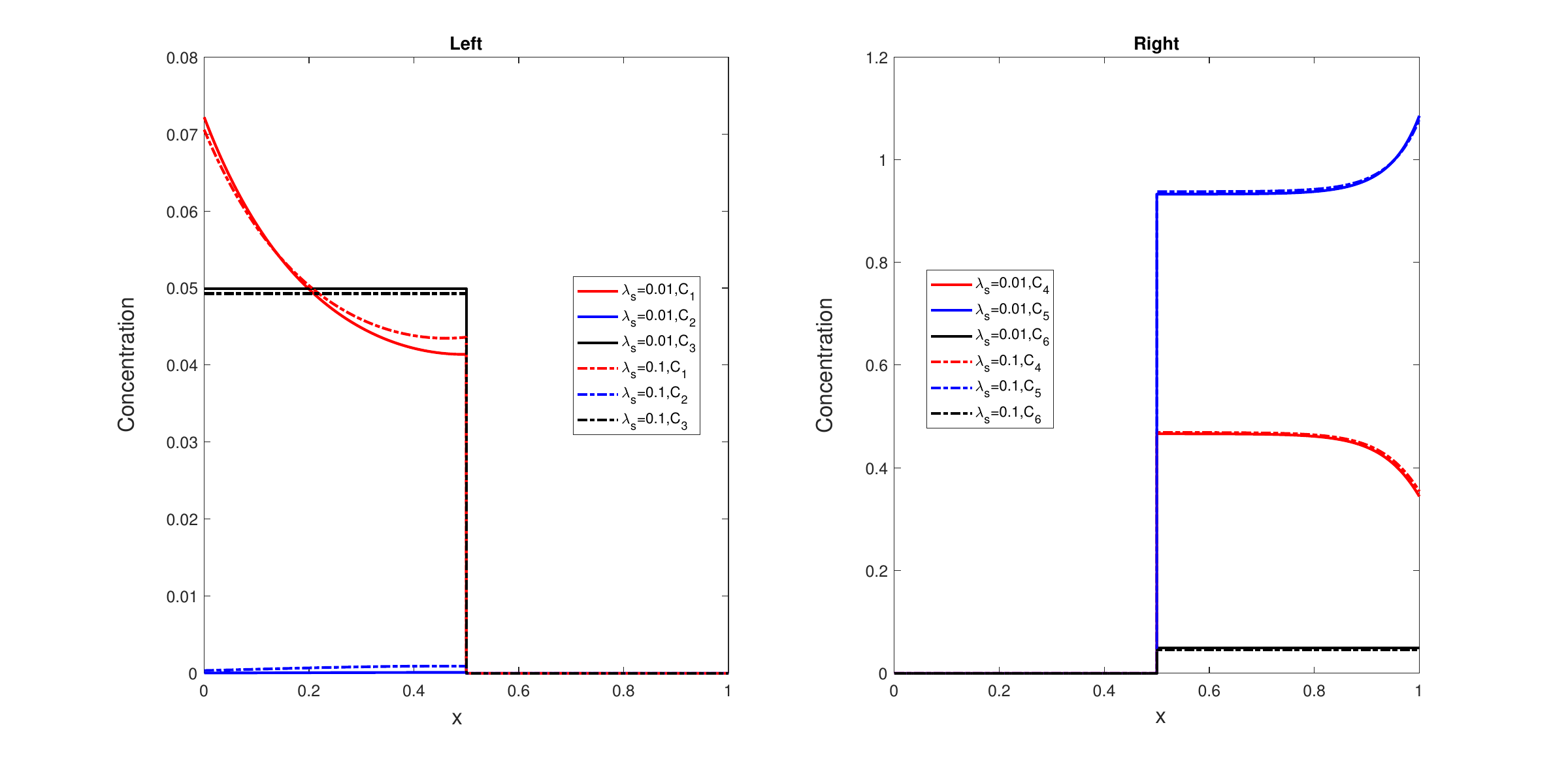}}
\end{subfigure}
\caption{  \textcolor{black}{Space distributions of    (a) Electric potential; (b) Substance concentrations at equilibrium with the same intrinsic reaction rate $k_{f,0}$ on both sides of the plate and different capacitance $\lambda_s$. Here $\zeta = 10$ is fixed. Dimensional units are used as defined in the beginning of the Results Section.}   }
\label{fig:lreqcm001space}
\end{figure}

In Fig. \ref{fig:lreqdiffcmtime} and Fig. \ref{fig:lreqcm001space}, we compare the results with different effective capacitance $\lambda_s$. The reaction rate increases as the capacitance increases. And more electrons could be stored on the capacitors. At the equilibrium, due to larger capacitance, the electric potential difference between the two sides electrolyte decreases. The effect on the  \textcolor{black}{spatial} distributions of substances at equilibrium is mainly on the left region due to the decrease of electric potential difference. 

\begin{figure}[htbp]
\centering
\begin{subfigure}[]{
		\includegraphics[width=0.25\textwidth]{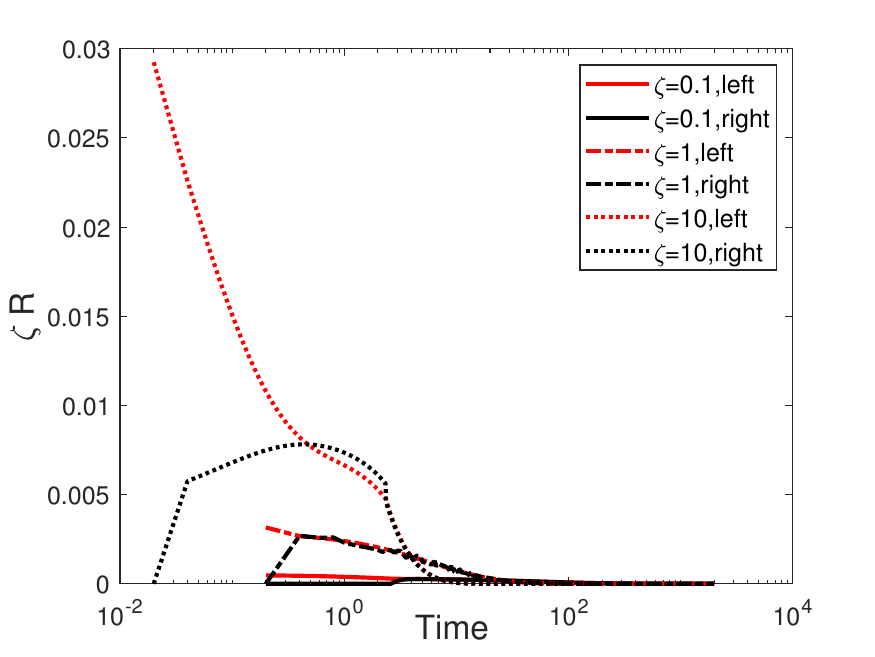}}
\end{subfigure}	
\begin{subfigure}[]{
		\includegraphics[width=0.25\textwidth]{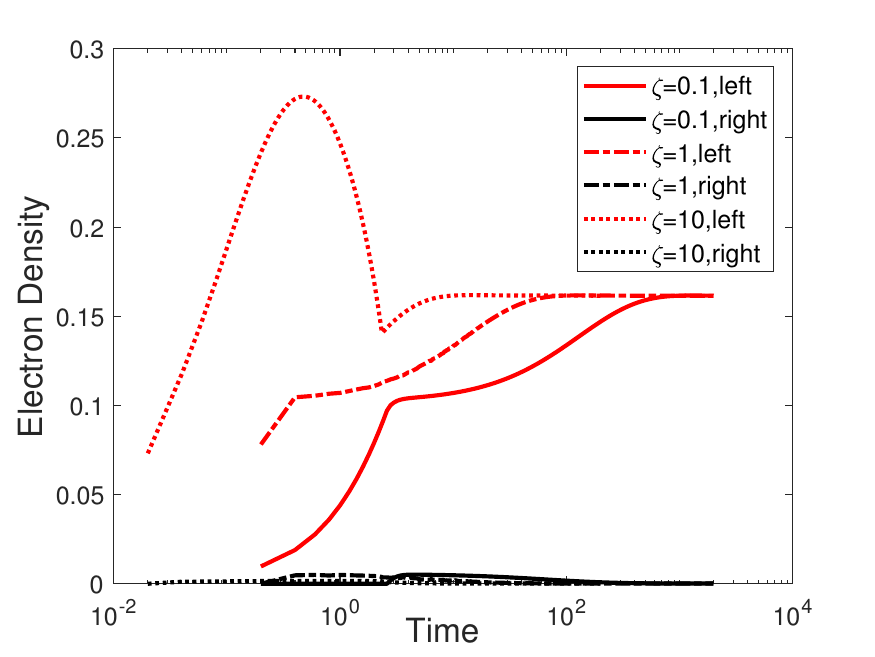}}
\end{subfigure}	

\begin{subfigure}[]{
		\includegraphics[width=0.45\textwidth]{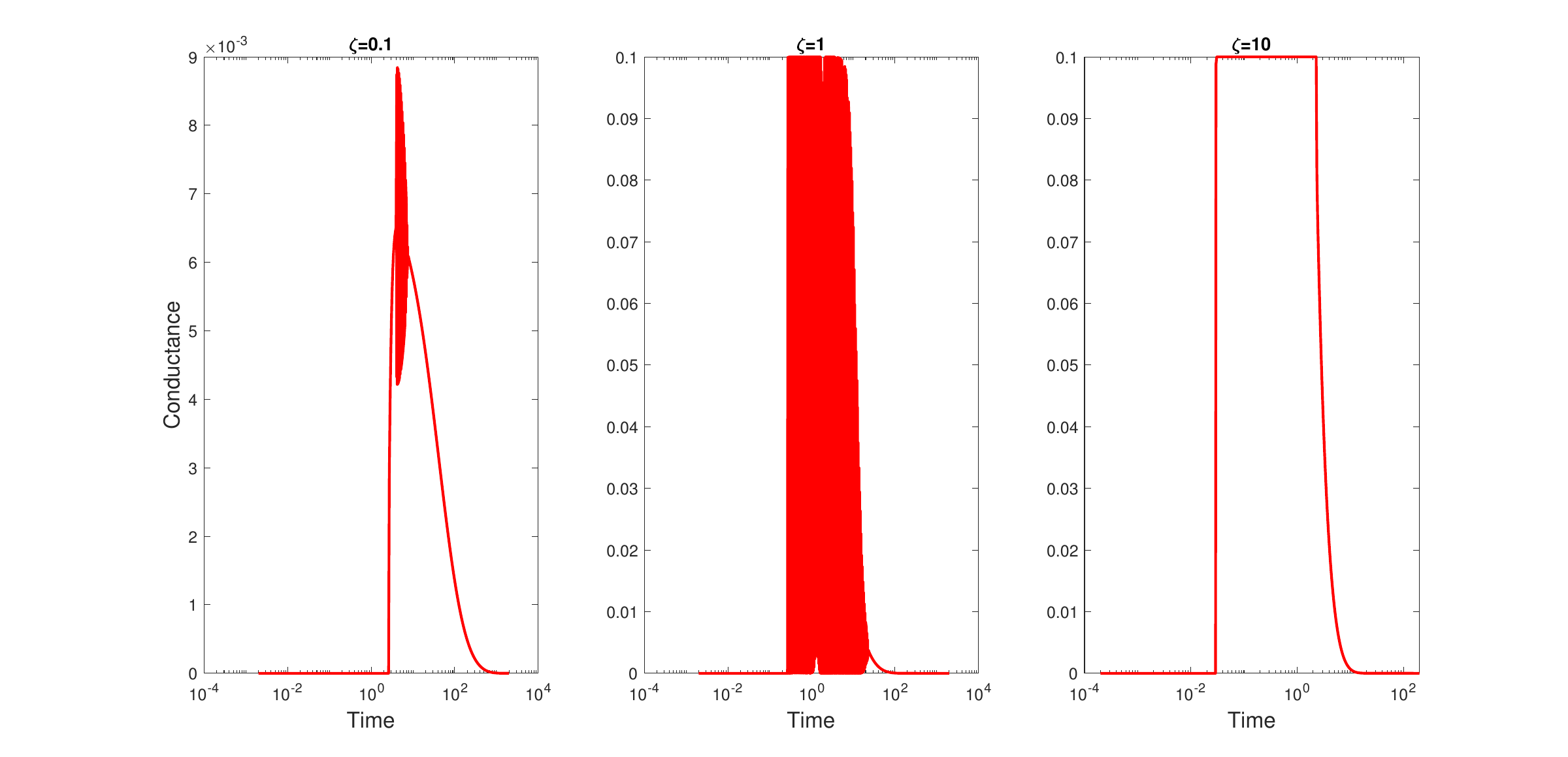}}
\end{subfigure}
\caption{  \textcolor{black}{Results with soft switch function.  Dynamics evolution of  (a)  Reaction rate;   (b) Electron density; (c) Conductance with fixed $\lambda_s=0.1$ and different timescale ratio $\zeta$.     Dimensional units are used as defined in the beginning of the Results Section.}}
\label{fig:lreqdiffcm_soft}
\end{figure}

 \textcolor{black}{Because the mechanism of the switch is not known, we investigate two types, a soft switch and a hard switch.  In Fig. \ref{fig:lreqdiffcm_soft}, we present the results obtained when utilizing the  \textcolor{black}{soft} switch function \eqref{softswitch} with $\epsilon = 0.001$. As shown in Fig. \ref{fig:lreqdiffcm_soft}a-b, a comparison with the hard switch reveals that the right reaction rate function $\mathcal{R}_r$ exhibits a smoother behavior, particularly noticeable for reaction-domainted case ($\zeta=10$).
This smoothing effect arises from the gradual transition around the threshold, allowing the switch to remain in an open state for an extended duration. Consequently, electrons can be transported continuously from the left to the right.}
 
\subsection{ \textcolor{black}{Different intrinsic reaction rates}}
\begin{figure}[!ht]
\centering
\begin{subfigure}[]{
		\includegraphics[width=0.35\textwidth]{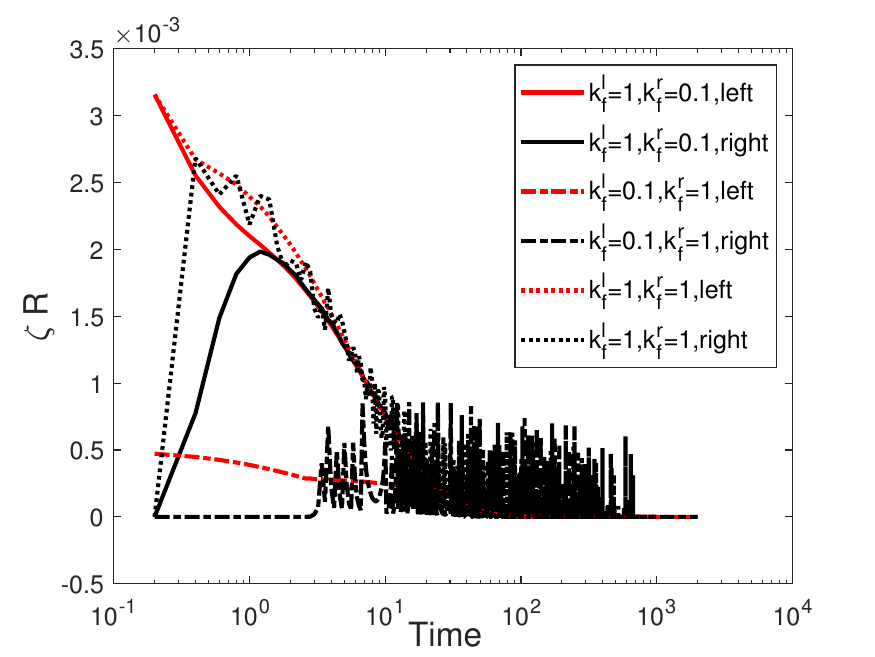}}
\end{subfigure}
\begin{subfigure}[]{
		\includegraphics[width=0.35\textwidth]{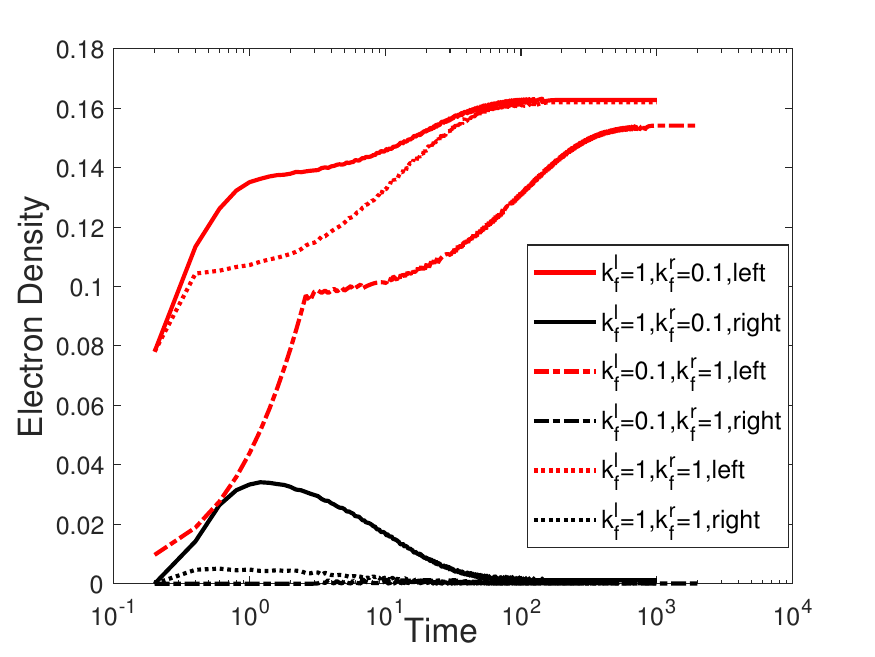}}
\end{subfigure}
\begin{subfigure}[]{
		\includegraphics[width=0.55\textwidth]{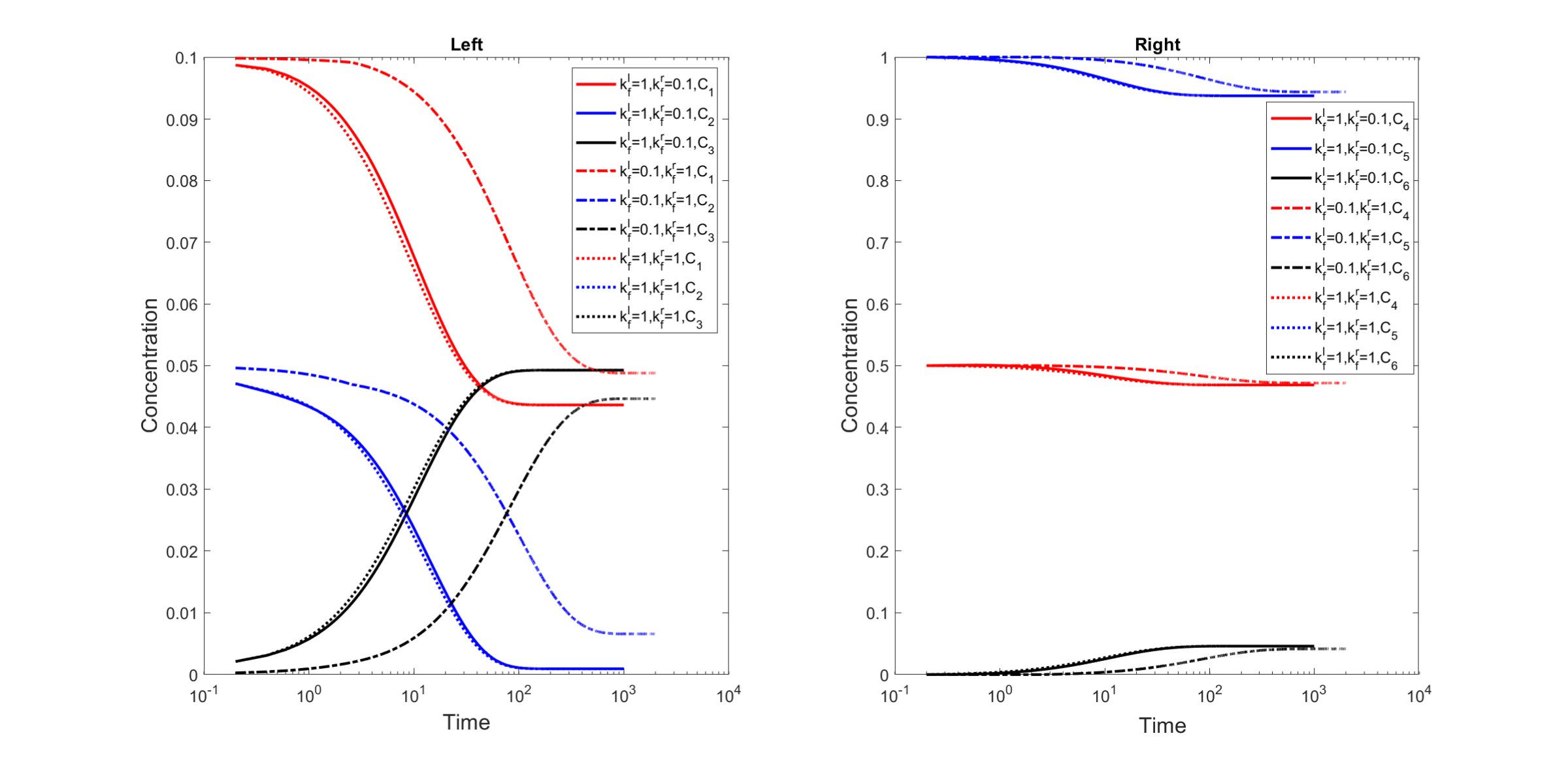}}
\end{subfigure}
\begin{subfigure}[]{
		\includegraphics[width=0.55\textwidth]{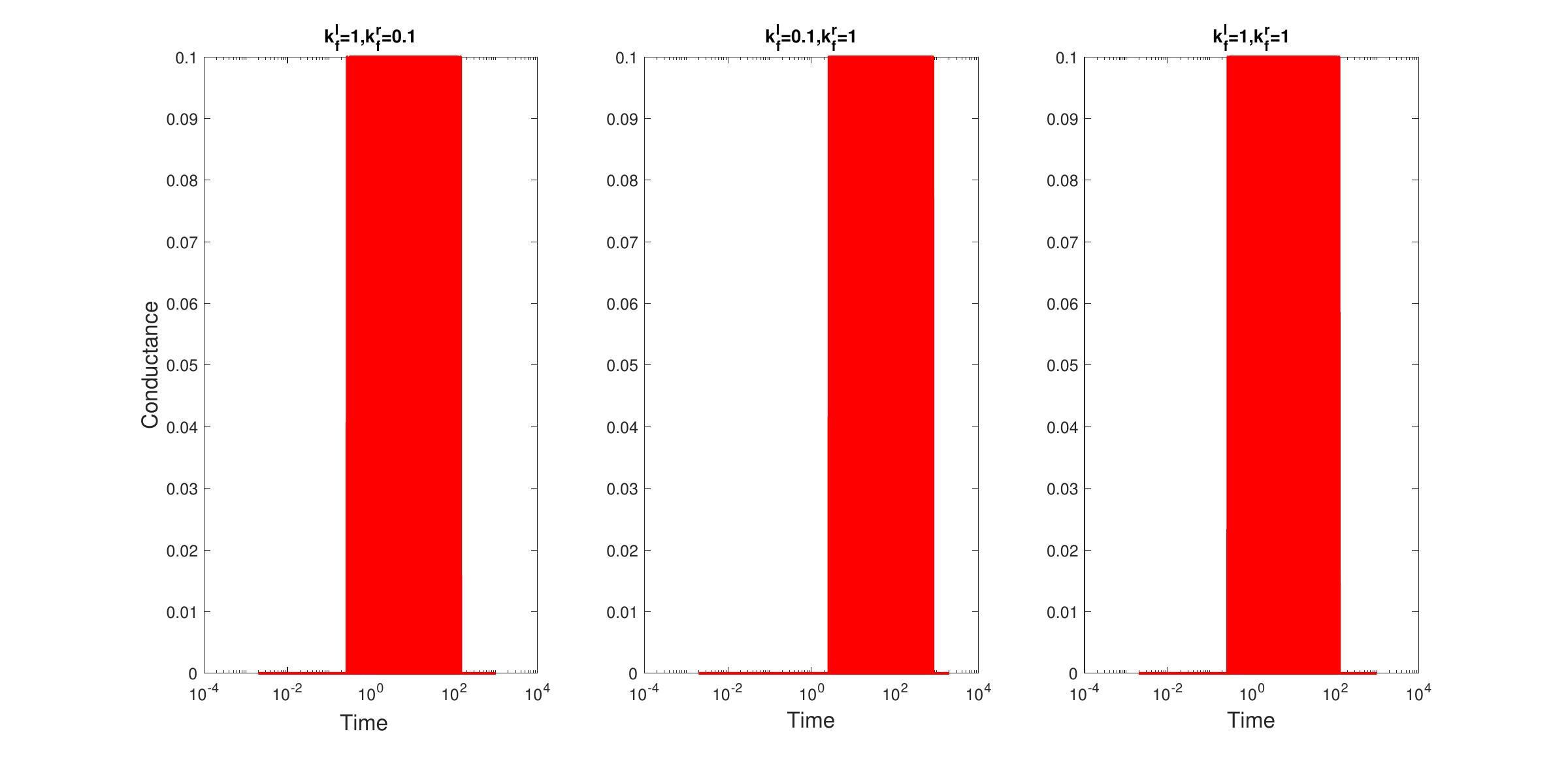}}
\end{subfigure}	

\caption{ \textcolor{black}{ Dynamics evolution of  (a) Reaction rates; (b) Electron Density; (c) Substance concentrations; (d) Conductance function over time with different intrinsic reation rates on both sides of the plate. Here we fix the effective capacitance $\lambda_s = 0.1$ and diffusion-reaction time scale ratio $\zeta = 1$. Hard switch function is used. Dimensional units are used as defined in the beginning of the Results Section.}  }
\label{fig:lrneqcm01time}
\end{figure}

 \textcolor{black}{In this subsection, we explore the  \textcolor{black}{more general} scenario where  \textcolor{black}{the intrinsic reaction rates on the left and right sides differ}, i.e., $k_{f,0}^l\neq k_{f,0}^r$. We maintain a fixed effective capacitance $\lambda_s = 0.1$ and a diffusion-reaction time scale ratio $\zeta = 1$. The outcomes are presented in Fig. \ref{fig:lrneqcm01time}, where solid lines represent the left-fast-right-slow case ($k_{f,0}^l=1, k_{f,0}^r=0.1$), dash-dot lines correspond to the left-slow-right-fast case ($k_{f,0}^l=0.1, k_{f,0}^r=1$), and dot lines represent the equal reaction case ($k_{f,0}^l=1, k_{f,0}^r=1$).}

 \textcolor{black}{When the left forward reaction rate constant is faster than the right one, electrons accumulate on the left plate more rapidly at the beginning. Consequently, the left plate reaction initially lags behind compared to the equal reaction case ($k_{f,0}^l=k_{f,0}^r=1$). The accumulated electrons activate the gate earlier than the slower accumulation (Fig. \ref{fig:lrneqcm01time}d). In cases where substances are rapidly depleted in the left region (Fig. \ref{fig:lrneqcm01time}c), the switch is closed. Since the right side forward reaction is slow ($k_{f,0}^r =0.1$), electrons accumulate on the right plates (Fig. \ref{fig:lrneqcm01time}d) and maintain the right reaction smoothly even after the gate is closed (Fig. \ref{fig:lrneqcm01time}a).}

 \textcolor{black}{In contrast, when the left forward reaction rate constant is slower than the right one, the electron density on the left plate increases more gradually. When the gate is open, the electrons are transported to the right \textcolor{red}{side}, initiating the reaction on the right later than in the other two cases (Fig. \ref{fig:lrneqcm01time}a). Simultaneously, since the right forward reaction rate is high, electrons are consumed rapidly, inducing high oscillations in the reaction function on the right plate and leading to high-frequency gating.}

\section{Discussion and Conclusion} \label{section:discussion}

This paper seeks to combine the classical description of  \textcolor{black}{electrochemical reactions with a general variational method.}  Based on the first and second laws of thermodynamics, we generalized the energy variation method \cite{shen2020energy,shen2022energy} for an open system with mass flux communication on the boundary.  Then the framework is used to study the chemical reaction in electrolytes where the charges are taken into consideration. When the reaction is  \textcolor{black}{only in the bulk region, and not at the membrane, the model obtained} is a Poisson-Nernst-Planck (PNP) system with reaction terms that could be taken as a generalized version of the diffusion-reaction model in \cite{wang2020field}. 

 \textcolor{black}{Then we consider the reactions on the boundary with current supplied from an external source. Note that our bidomain model (with different reactions on each side of the membrane) forms a self-regulatory reaction system.}  The effect of electric potential on the reaction rate is consistently  \textcolor{black}{modeled with a version of the Frumkin-Butler-Volmer Equation }\cite{biesheuvel2009imposed,sekimoto2010stochastic,van2010diffuse,van2012frumkin,fraggedakis2021theory}.  

Next, we  \textcolor{black}{analyze a bidomain model with (1) oxidation on the left input side that produces electrons and (2) reduction on the right output side that consumes electrons. Transport across the membrane of conduction and displacement current is included.}  

 \textcolor{black}{We analyze this system because it plays an important role in ATP production and the transport of oxygen/carbon dioxide in living systems. Serious disruption in these reactions is likely to be lethal. Small disruptions lead to a range of health problems, including anemia, sickle cell disease, and other blood disorders.}

The simulation results are used to illustrate the diffusion-reaction competition and how the electric potential affects the reaction rate function. In order to model the gating phenomena, we presented simple both hard switch and soft switch models that assume the conductivity of the interface depends on the electric potential difference like the Hodgkin-Huxley model \cite{hodgkin1952measurement,hodgkin1949ionic}.

Our treatment, however, uses only the simplest representations of the chemical reactions themselves and we are quite aware that the specifics of the reactions that have been so exhaustively analyzed in the electrochemistry literature will need to be introduced to our current based models as more realism is sought.  \textcolor{black}{Different systems in electrochemistry have different mechanisms. Different systems in biology are also likely to use somewhat different mechanisms for the reactions and the switch, with different regulation.}

In this regard, it will be of particular interest to biologists to consult the literature on proton-electron reactions in inorganic systems \cite{fraggedakis2021theory}. In the future, with more detailed information on the reactions and biological structure,  \textcolor{black}{our model could be used to study the full electron transport chain during the ATP production that occurs in almost every cell in an animal }and other processes, perhaps including the blood clotting cascade process \cite{xu2010multiscale,smith2015all}.

Chemical reactions can also play an important role in the active motion of soft matter. For example, in some active gels, chemical reactions between the gel components can generate mechanical stress that drives the motion of the gel. Similarly, chemical reactions can be used to generate motion in droplets or other active soft matter systems.
Also, in the current work, the fluid is neglected temporarily for simplicity. By adding the kinetic energy of the fluid to the total energy $E$ and constitutive relation,  our model could be used to study the active rheology of soft matter  \cite{wang2021onsager}.

\section*{Acknowledgments}
This work was partially supported by the 
National Natural Science Foundation of China  no. 12071190, 12231004,  and Natural Sciences and Engineering Research Council of Canada (NSERC).
S. Xu   thanks Prof. Kai Zhang and Prof. Changcheng Zheng for suggestions and helpful discussions.

\newpage
\bibliography{mybib_new}

\begin{thebibliography}{10}

\bibitem{bazant2012theory}
Martin~Z Bazant.
\newblock Theory of electrochemical kinetics based on nonequilibrium
  thermodynamics, 2012.

\bibitem{bazant2013theory}
Martin~Z Bazant.
\newblock Theory of chemical kinetics and charge transfer based on
  nonequilibrium thermodynamics.
\newblock {\em Accounts of chemical research}, 46(5):1144--1160, 2013.

\bibitem{bazant2005current}
Martin~Z Bazant, Kevin~T Chu, and Bruce~J Bayly.
\newblock Current-voltage relations for electrochemical thin films.
\newblock {\em SIAM journal on applied mathematics}, 65(5):1463--1484, 2005.

\bibitem{belevich2007exploring}
Ilya Belevich, Dmitry~A Bloch, Nikolai Belevich, M{\aa}rten Wikstr{\"o}m, and
  Michael~I Verkhovsky.
\newblock Exploring the proton pump mechanism of cytochrome c oxidase in real
  time.
\newblock {\em Proceedings of the National Academy of Sciences},
  104(8):2685--2690, 2007.

\bibitem{bezanilla2006action}
Francisco Bezanilla.
\newblock The action potential: from voltage-gated conductances to molecular
  structures.
\newblock {\em Biological research}, 39(3):425--435, 2006.

\bibitem{biesheuvel2009imposed}
PM~Biesheuvel, Michiel Van~Soestbergen, and Martin~Z Bazant.
\newblock Imposed currents in galvanic cells.
\newblock {\em Electrochimica Acta}, 54(21):4857--4871, 2009.

\bibitem{bloch2004catalytic}
Dmitry Bloch, Ilya Belevich, Audrius Jasaitis, Camilla Ribacka, Anne Puustinen,
  Michael~I Verkhovsky, and M{\aa}rten Wikstr{\"o}m.
\newblock The catalytic cycle of cytochrome c oxidase is not the sum of its two
  halves.
\newblock {\em Proceedings of the National Academy of Sciences},
  101(2):529--533, 2004.

\bibitem{blomberg2012mechanism}
Margareta~RA Blomberg and Per~EM Siegbahn.
\newblock The mechanism for proton pumping in cytochrome c oxidase from an
  electrostatic and quantum chemical perspective.
\newblock {\em Biochimica et Biophysica Acta (BBA)-Bioenergetics},
  1817(4):495--505, 2012.

\bibitem{bonnefont2001analysis}
Antoine Bonnefont, Fran{\c{c}}oise Argoul, and Martin~Z Bazant.
\newblock Analysis of diffuse-layer effects on time-dependent interfacial
  kinetics.
\newblock {\em Journal of Electroanalytical Chemistry}, 500(1-2):52--61, 2001.

\bibitem{brunet2004generalized}
Edouard Brunet and Armand Ajdari.
\newblock Generalized onsager relations for electrokinetic effects in
  anisotropic and heterogeneous geometries.
\newblock {\em Physical Review E}, 69(1):016306, 2004.

\bibitem{brush1976kind}
Stephen~G Brush.
\newblock {\em The kind of motion we call heat}, volume~2.
\newblock North-Holland Amsterdam, 1976.

\bibitem{butler1924studies}
J~A\_~V Butler.
\newblock Studies in heterogeneous equilibria. part ii.—the kinetic
  interpretation of the nernst theory of electromotive force.
\newblock {\em Transactions of the Faraday Society}, 19(March):729--733, 1924.

\bibitem{cai2018network}
Xiuhong Cai, Kamran Haider, Jianxun Lu, Slaven Radic, Chang~Yun Son, Qiang Cui,
  and MR~Gunner.
\newblock Network analysis of a proposed exit pathway for protons to the p-side
  of cytochrome c oxidase.
\newblock {\em Biochimica et Biophysica Acta (BBA)-Bioenergetics},
  1859(10):997--1005, 2018.

\bibitem{catterall2000ionic}
William~A Catterall.
\newblock From ionic currents to molecular mechanisms: the structure and
  function of voltage-gated sodium channels.
\newblock {\em Neuron}, 26(1):13--25, 2000.

\bibitem{chree1908mathematical}
C~Chree.
\newblock The mathematical theory of electricity and magnetism.
\newblock {\em Nature}, 78(2031):537--538, 1908.

\bibitem{de2019reaction}
Kevin~C de~Berg and Kevin~C de~Berg.
\newblock The reaction: Chemical affinity: Laws, theories and models.
\newblock {\em The Iron (III) Thiocyanate Reaction: Research History and Role
  in Chemical Analysis}, pages 41--52, 2019.

\bibitem{de2019traite}
Antoine-Laurent de~Lavoisier.
\newblock {\em Trait{\'e} {\'e}l{\'e}mentaire de chimie}.
\newblock Maxtor France, 2019.

\bibitem{dickinson2020butler}
Edmund~JF Dickinson and Andrew~J Wain.
\newblock The butler-volmer equation in electrochemical theory: Origins, value,
  and practical application.
\newblock {\em Journal of Electroanalytical Chemistry}, 872:114145, 2020.

\bibitem{doi1983variational}
Masao Doi.
\newblock Variational principle for the kirkwood theory for the dynamics of
  polymer solutions and suspensions.
\newblock {\em The Journal of chemical physics}, 79(10):5080--5087, 1983.

\bibitem{edwards1974theory}
SF~Edwards and Karl~F Freed.
\newblock Theory of the dynamical viscosity of polymer solutions.
\newblock {\em The Journal of Chemical Physics}, 61(3):1189--1202, 1974.

\bibitem{eisenberg2010energy}
Bob Eisenberg, Yunkyong Hyon, and Chun Liu.
\newblock Energy variational analysis of ions in water and channels: Field
  theory for primitive models of complex ionic fluids.
\newblock {\em The Journal of Chemical Physics}, 133(10):104104, 2010.

\bibitem{erdey1930theorie}
Tibor Erdey-Gr{\'u}z and Max Volmer.
\newblock Zur theorie der wasserstoff {\"u}berspannung.
\newblock {\em Zeitschrift f{\"u}r physikalische Chemie}, 150(1):203--213,
  1930.

\bibitem{feynman2006feynman}
RP0138 Feynman, RB~Leighton, and ML~Sands.
\newblock Feynman lectures on physics, vol. 2, lecture 19, 2006.

\bibitem{finlayson1972existence}
Bruce~A Finlayson.
\newblock Existence of variational principles for the navier-stokes equation.
\newblock {\em The physics of fluids}, 15(6):963--967, 1972.

\bibitem{fraggedakis2021theory}
Dimitrios Fraggedakis, Michael McEldrew, Raymond~B Smith, Yamini Krishnan,
  Yirui Zhang, Peng Bai, William~C Chueh, Yang Shao-Horn, and Martin~Z Bazant.
\newblock Theory of coupled ion-electron transfer kinetics.
\newblock {\em Electrochimica Acta}, 367:137432, 2021.

\bibitem{gadsby2009ion}
David~C Gadsby.
\newblock Ion channels versus ion pumps: the principal difference, in
  principle.
\newblock {\em Nature reviews Molecular cell biology}, 10(5):344--352, 2009.

\bibitem{garber1995maxwell}
Elizabeth Garber, Stephen~G Brush, and CWF Everitt.
\newblock {\em Maxwell on Heat and Statistical Mechanics: On" Avoiding All
  Personal Enquiries" of Molecules}.
\newblock Lehigh University Press, 1995.

\bibitem{ge2016mesoscopic}
Hao Ge and Hong Qian.
\newblock Mesoscopic kinetic basis of macroscopic chemical thermodynamics: A
  mathematical theory.
\newblock {\em Physical Review E}, 94(5):052150, 2016.

\bibitem{gennes2004capillarity}
Pierre-Gilles Gennes, Fran{\c{c}}oise Brochard-Wyart, David Qu{\'e}r{\'e},
  et~al.
\newblock {\em Capillarity and wetting phenomena: drops, bubbles, pearls,
  waves}.
\newblock Springer, 2004.

\bibitem{guo2018structure}
Runyu Guo, Jinke Gu, Shuai Zong, Meng Wu, and Maojun Yang.
\newblock Structure and mechanism of mitochondrial electron transport chain.
\newblock {\em Biomedical journal}, 41(1):9--20, 2018.

\bibitem{hodgkin1949ionic}
AL~im HODGKIN.
\newblock Ionic currents underlying activity in the giant axon of the squid.
\newblock {\em Arch. Sci. Physiol.}, 3:129--150, 1949.

\bibitem{hodgkin1952quantitative}
Alan~L Hodgkin and Andrew~F Huxley.
\newblock A quantitative description of membrane current and its application to
  conduction and excitation in nerve.
\newblock {\em The Journal of physiology}, 117(4):500, 1952.

\bibitem{hodgkin1952measurement}
Alan~L Hodgkin, Andrew~F Huxley, and Bernard Katz.
\newblock Measurement of current-voltage relations in the membrane of the giant
  axon of loligo.
\newblock {\em The Journal of physiology}, 116(4):424, 1952.

\bibitem{knopf2021phase}
Patrik Knopf, Kei~Fong Lam, Chun Liu, and Stefan Metzger.
\newblock Phase-field dynamics with transfer of materials: the cahn--hilliard
  equation with reaction rate dependent dynamic boundary conditions.
\newblock {\em ESAIM: Mathematical Modelling and Numerical Analysis},
  55(1):229--282, 2021.

\bibitem{liu2012hydrodynamic}
Chun Liu, Tiezheng Qian, and Xinpeng Xu.
\newblock Hydrodynamic boundary conditions for one-component liquid-gas flows
  on non-isothermal solid substrates.
\newblock {\em Communications in Mathematical Sciences}, 10(4):1027--1053,
  2012.

\bibitem{liu2021structure}
Chun Liu, Cheng Wang, and Yiwei Wang.
\newblock A structure-preserving, operator splitting scheme for
  reaction-diffusion equations with detailed balance.
\newblock {\em Journal of Computational Physics}, 436:110253, 2021.

\bibitem{liu2019energetic}
Chun Liu and Hao Wu.
\newblock An energetic variational approach for the cahn--hilliard equation
  with dynamic boundary condition: model derivation and mathematical analysis.
\newblock {\em Archive for Rational Mechanics and Analysis}, 233(1):167--247,
  2019.

\bibitem{mielke2017non}
Alexander Mielke, Robert~IA Patterson, Mark~A Peletier, and DR3691731
  Michiel~Renger.
\newblock Non-equilibrium thermodynamical principles for chemical reactions
  with mass-action kinetics.
\newblock {\em SIAM Journal on Applied Mathematics}, 77(4):1562--1585, 2017.

\bibitem{moya1995ionic}
AA~Moya, J~Castilla, and J~Horno.
\newblock Ionic transport in electrochemical cells including electrical
  double-layer effects. a network thermodynamics approach.
\newblock {\em The Journal of Physical Chemistry}, 99(4):1292--1298, 1995.

\bibitem{murphy1992numerical}
WD~Murphy, JA~Manzanares, S~Maf{\'e}, and H~Reiss.
\newblock A numerical study of the equilibrium and nonequilibrium diffuse
  double layer in electrochemical cells.
\newblock {\em The Journal of Physical Chemistry}, 96(24):9983--9991, 1992.

\bibitem{nelson2008lehninger}
David~L Nelson, Albert~L Lehninger, and Michael~M Cox.
\newblock {\em Lehninger principles of biochemistry}.
\newblock Macmillan, 2008.

\bibitem{nieves2018regulation}
Madeline Nieves-Cintr{\'o}n, Arsalan~U Syed, Matthew~A Nystoriak, and Manuel~F
  Navedo.
\newblock Regulation of voltage-gated potassium channels in vascular smooth
  muscle during hypertension and metabolic disorders.
\newblock {\em Microcirculation}, 25(1):e12423, 2018.

\bibitem{onsager1931reciprocal1}
Lars Onsager.
\newblock Reciprocal relations in irreversible processes. i.
\newblock {\em Physical review}, 37(4):405, 1931.

\bibitem{onsager1931reciprocal2}
Lars Onsager.
\newblock Reciprocal relations in irreversible processes. ii.
\newblock {\em Physical review}, 38(12):2265, 1931.

\bibitem{ozcan2022equilibrium}
Muzaffer ozcan.
\newblock Why equilibrium constants are unitless.
\newblock {\em The Journal of Physical Chemistry Letters}, 13(15):3507--3509,
  2022.

\bibitem{qian2006variational}
Tiezheng Qian, Xiao-Ping Wang, and Ping Sheng.
\newblock A variational approach to moving contact line hydrodynamics.
\newblock {\em Journal of Fluid Mechanics}, 564:333--360, 2006.

\bibitem{ren2010continuum}
Weiqing Ren, Dan Hu, and Weinan E.
\newblock Continuum models for the contact line problem.
\newblock {\em Physics of fluids}, 22(10):102103, 2010.

\bibitem{ryham2006energetic}
Rolf~Josef Ryham.
\newblock An energetic variational approach to mathematical modeling of charged
  fluids: charge phases, simulation and well posedness.
\newblock 2006.

\bibitem{sekimoto2010stochastic}
Ken Sekimoto.
\newblock {\em Stochastic energetics}, volume 799.
\newblock Springer, 2010.

\bibitem{shen2020energy}
Lingyue Shen, Huaxiong Huang, Ping Lin, Zilong Song, and Shixin Xu.
\newblock An energy stable c0 finite element scheme for a quasi-incompressible
  phase-field model of moving contact line with variable density.
\newblock {\em Journal of Computational Physics}, 405:109179, 2020.

\bibitem{shen2022energy}
Lingyue Shen, Zhiliang Xu, Ping Lin, Huaxiong Huang, and Shixin Xu.
\newblock An energy stable c\^{}0 finite element scheme for a phase-field model
  of vesicle motion and deformation.
\newblock {\em SIAM Journal on Scientific Computing}, 44(1):B122--B145, 2022.

\bibitem{simpson1998maxwell}
Thomas~K Simpson.
\newblock Maxwell on the electromagnetic field: A guided study., 1998.

\bibitem{smith2015all}
Stephanie~A Smith, Richard~J Travers, and James~H Morrissey.
\newblock How it all starts: Initiation of the clotting cascade.
\newblock {\em Critical reviews in biochemistry and molecular biology},
  50(4):326--336, 2015.

\bibitem{strutt1871some}
JW~Strutt.
\newblock Some general theorems relating to vibrations.
\newblock {\em Proceedings of the London Mathematical Society}, 1(1):357--368,
  1871.

\bibitem{truesdell1969rational}
Clifford Truesdell.
\newblock {\em Rational thermodynamics: a course of lectures on selected
  topics}.
\newblock McGraw-Hill, 1969.

\bibitem{van2010diffuse}
M~Van~Soestbergen, PM~Biesheuvel, and Martin~Z Bazant.
\newblock Diffuse-charge effects on the transient response of electrochemical
  cells.
\newblock {\em Physical Review E}, 81(2):021503, 2010.

\bibitem{van2012frumkin}
Michiel Van~Soestbergen.
\newblock Frumkin-butler-volmer theory and mass transfer in electrochemical
  cells.
\newblock {\em Russian journal of electrochemistry}, 48(6):570--579, 2012.

\bibitem{verkhovskaya2008real}
Marina~L Verkhovskaya, Nikolai Belevich, Liliya Euro, M{\aa}rten Wikstr{\"o}m,
  and Michael~I Verkhovsky.
\newblock Real-time electron transfer in respiratory complex i.
\newblock {\em Proceedings of the National Academy of Sciences},
  105(10):3763--3767, 2008.

\bibitem{verkhovsky2006elementary}
Michael~I Verkhovsky, Ilya Belevich, Dmitry~A Bloch, and M{\aa}rten
  Wikstr{\"o}m.
\newblock Elementary steps of proton translocation in the catalytic cycle of
  cytochrome oxidase.
\newblock {\em Biochimica et Biophysica Acta (BBA)-Bioenergetics},
  1757(5-6):401--407, 2006.

\bibitem{wang2021onsager}
Haiqin Wang, Tiezheng Qian, and Xinpeng Xu.
\newblock Onsager's variational principle in active soft matter.
\newblock {\em Soft Matter}, 17(13):3634--3653, 2021.

\bibitem{wang2021generalized}
Qi~Wang.
\newblock Generalized onsager principle and it applications.
\newblock {\em Frontiers and Progress of Current Soft Matter Research}, pages
  101--132, 2021.

\bibitem{wang2022some}
Yiwei Wang and Chun Liu.
\newblock Some recent advances in energetic variational approaches.
\newblock {\em Entropy}, 24(5):721, 2022.

\bibitem{wang2020field}
Yiwei Wang, Chun Liu, Pei Liu, and Bob Eisenberg.
\newblock Field theory of reaction-diffusion: Law of mass action with an
  energetic variational approach.
\newblock {\em Physical Review E}, 102(6):062147, 2020.

\bibitem{wikstrom2018proton}
M{\aa}rten Wikstr{\"o}m and Vivek Sharma.
\newblock Proton pumping by cytochrome c oxidase--a 40 year anniversary.
\newblock {\em Biochimica et Biophysica Acta (BBA)-Bioenergetics},
  1859(9):692--698, 2018.

\bibitem{wikstrom2003water}
M{\aa}rten Wikstr{\"o}m, Michael~I Verkhovsky, and Gerhard Hummer.
\newblock Water-gated mechanism of proton translocation by cytochrome c
  oxidase.
\newblock {\em Biochimica et Biophysica Acta (BBA)-Bioenergetics},
  1604(2):61--65, 2003.

\bibitem{xu2019numerical}
Shixin Xu, Xinfu Chen, Chun Liu, and Xingye Yue.
\newblock Numerical method for multi-alleles genetic drift problem.
\newblock {\em SIAM Journal on Numerical Analysis}, 57(4):1770--1788, 2019.

\bibitem{xu2014energetic}
Shixin Xu, Ping Sheng, and Chun Liu.
\newblock An energetic variational approach for ion transport.
\newblock {\em arXiv preprint arXiv:1408.4114}, 2014.

\bibitem{xu2019generalized}
Xinpeng Xu and Tiezheng Qian.
\newblock Generalized lorentz reciprocal theorem in complex fluids and in
  non-isothermal systems.
\newblock {\em Journal of Physics: Condensed Matter}, 31(47):475101, 2019.

\bibitem{xu2010multiscale}
Zhiliang Xu, Joshua Lioi, Jian Mu, Malgorzata~M Kamocka, Xiaomin Liu, Danny~Z
  Chen, Elliot~D Rosen, and Mark Alber.
\newblock A multiscale model of venous thrombus formation with surface-mediated
  control of blood coagulation cascade.
\newblock {\em Biophysical journal}, 98(9):1723--1732, 2010.

\bibitem{yan2021adaptive}
David Yan, Mary~C Pugh, and Francis~P Dawson.
\newblock Adaptive time-stepping schemes for the solution of the
  poisson-nernst-planck equations.
\newblock {\em Applied Numerical Mathematics}, 163:254--269, 2021.

\bibitem{yang2016hydrodynamic}
Xiaogang Yang, Jun Li, M~Gregory Forest, and Qi~Wang.
\newblock Hydrodynamic theories for flows of active liquid crystals and the
  generalized onsager principle.
\newblock {\em Entropy}, 18(6):202, 2016.

\bibitem{yong2012conservation}
Wen-An Yong.
\newblock Conservation-dissipation structure of chemical reaction systems.
\newblock {\em Physical Review E}, 86(6):067101, 2012.

\bibitem{zangwill2013modern}
Andrew Zangwill.
\newblock {\em Modern electrodynamics}.
\newblock Cambridge University Press, 2013.

\bibitem{zhang2014derivation}
Zhen Zhang, Shixin Xu, and Weiqing Ren.
\newblock Derivation of a continuum model and the energy law for moving contact
  lines with insoluble surfactants.
\newblock {\em Physics of Fluids}, 26(6):062103, 2014.

\bibitem{zhao2021thermodynamically}
Quan Zhao, Weiqing Ren, and Zhen Zhang.
\newblock A thermodynamically consistent model and its conservative numerical
  approximation for moving contact lines with soluble surfactants.
\newblock {\em Computer Methods in Applied Mechanics and Engineering},
  385:114033, 2021.

\bibitem{zhao2019mitochondrial}
Ru-Zhou Zhao, Shuai Jiang, Lin Zhang, and Zhi-Bin Yu.
\newblock Mitochondrial electron transport chain, ros generation and
  uncoupling.
\newblock {\em International journal of molecular medicine}, 44(1):3--15, 2019.

\end{thebibliography}

\newpage
\appendix

\section{ \textcolor{black}{Details of derivation}}

\noindent\textbf{Eq.\eqref{eqn: dedtingeneral} derivation}
\begin{eqnarray}
	\frac{dE}{dt} &=&\int_{\Omega}\sum_{i=1}^N\left\{\mu_i \frac {\partial C_i}{\partial t}\right\}dx +\int_{\Omega} \bE\cdot\frac{\partial \bD}{\partial t}dx\nonumber\\
	&=&\int_{\Omega}\sum_{i=1}^N\left\{\mu_i \frac {\partial C_i}{\partial t}\right\}dx -\int_{\Omega}\nabla \phi\cdot\frac{\partial\bD}{\partial t}dx\nonumber\\
	&=&  \int_{\Omega}\sum_{i=1}^N\left\{\mu_i \frac {\partial C_i}{\partial t}\right\}dx+\int_{\Omega} \phi\nabla\cdot\left(\frac{\partial \bD}{\partial t}\right)dx\nonumber\\
	&=&  \int_{\Omega}\sum_{i=1}^N\left\{\mu_i \frac {\partial C_i}{ \partial t}\right\}dx +\int_{\Omega} \phi F\sum_{i=1}^N\left\{z_i \frac {\partial C_i}{\partial t}\right\} dx\nonumber\\
	&=&  \int_{\Omega} \sum_{i=1}^N\left\{\tilde{\mu}_i \frac {\partial C_i}{\partial t}\right\}dx\nonumber\\
	&=&-\int_{\Omega}\sum_{i=1}^N\left\{\tilde{\mu}_i\nabla\cdot \boldsymbol{j}_i\right\}dx
	+\int_{\Omega}\mathcal{R}\sum_{i=1}^N\gamma_i  \tilde{\mu}_i dx\nonumber \\
	&=&\int_{\Omega}\sum_{i=1}^N\left\{\nabla\tilde{\mu}_i\cdot \boldsymbol{j}_i\right\}dx +\int_{\Omega}\mathcal{R} \sum_{i=1}^N\gamma_i  \tilde{\mu}_i dx-\int_{\partial\Omega} \sum_{i=1}^N (\tilde{\mu}_i-\tilde{\mu}_{i,ex}) \bj_i\cdot\boldsymbol{n} dS -\int_{\partial\Omega} \tilde{\mu}_{i,ex} \bj_i\cdot\boldsymbol{n} dS.\nonumber \\
	&=&-\Delta +\mathcal{P}_{E,\partial\Omega}.\nonumber
\end{eqnarray}

\noindent\textbf{Eq.\eqref{eqn:dedtboundary} derivation}
\begin{eqnarray} 
&&\frac{dE}{dt}\nonumber\\
&=& \int_{\Omega}\sum_{i=1}^N \left\{ \tilde{\mu}_i\frac{\partial C_i}{\partial t}\right\} dx-\int_{\partial \Omega}\phi\frac{\partial \boldsymbol{D}\cdot\boldsymbol{n}}{\partial t}dS +\int_{\Gamma}\mu_e\frac{\partial C_e}{\partial t}dS +\int_{\Gamma}(\phi-\phi_p) F \frac{\partial C_e}{\partial t} dS\nonumber\\
&=& -\int_{\Omega}\sum_{i=1}^N \left\{ \tilde{\mu}_i\nabla\cdot \boldsymbol{j}_i\right\} dx-\int_{\partial \Omega}\phi\frac{\partial \boldsymbol{D}\cdot\boldsymbol{n}}{\partial t}dS +\int_{\Gamma}\tilde{\mu}_e\frac{\partial C_e}{\partial t}dS +\int_{\Gamma}\phi F \frac{\partial C_e}{\partial t} dS\nonumber\\
&=& -\int_{\Omega}\sum_{i=1}^N \left\{ \tilde{\mu}_i\nabla\cdot \boldsymbol{j}_i\right\} dx-\int_{\Gamma}\phi\frac{\partial  }{\partial t}\left(\boldsymbol{D}\cdot\boldsymbol{n} - FC_e\right)dS+\int_{\Gamma}\tilde{\mu}_e (j_{ex}-\Delta z \mathcal{R})dS\nonumber\\
&&-\int_{\partial\Omega/\Gamma}\phi\frac{\partial \boldsymbol{D}\cdot\boldsymbol{n}}{\partial t}dS\nonumber\\
&=& \int_{\Omega}\sum_{i=1}^N \left\{ \nabla\tilde{\mu}_i\cdot \boldsymbol{j}_i\right\} dx-\int_{\Gamma}\phi\frac{\partial  }{\partial t}\left(\boldsymbol{D}\cdot\boldsymbol{n} - FC_e\right)dS-\int_{\Gamma}\sum_i^N\tilde{\mu}_i \boldsymbol{j}_i\cdot\boldsymbol{n}dS +\int_{\Gamma}\tilde{\mu}_e (-\Delta z \mathcal{R})dS\nonumber\\
&&-\int_{\partial\Omega/\Gamma}\phi\frac{\partial \boldsymbol{D}\cdot\boldsymbol{n}}{\partial t}dS-\int_{\partial\Omega/\Gamma}\sum_i^N\tilde{\mu}_i \boldsymbol{j}_i\cdot\boldsymbol{n}dS+\int_{\Gamma}\tilde{\mu}_e j_{ex}dS\nonumber\\
&=& \int_{\Omega}\sum_{i=1}^N \left\{ \nabla\tilde{\mu}_i\cdot \boldsymbol{j}_i\right\} dx-\int_{\Gamma}\phi\frac{\partial  }{\partial t}\left(\boldsymbol{D}\cdot\boldsymbol{n} - FC_e\right)dS+\int_{\Gamma}\sum_i^N\tilde{\mu}_i \gamma_i\mathcal{R}dS -\int_{\Gamma}\tilde{\mu}_e (\Delta z \mathcal{R})dS\nonumber\\
&&-\int_{\partial\Omega/\Gamma}\phi\frac{\partial \boldsymbol{D}\cdot\boldsymbol{n}}{\partial t}dS-\int_{\partial\Omega/\Gamma}\sum_i^N\tilde{\mu}_i \boldsymbol{j}_i\cdot\boldsymbol{n}dS +\int_{\Gamma}(\tilde{\mu}_e -\tilde{\mu}_{ex})j_{ex}dS+\int_{\Gamma}\tilde{\mu}_{ex} j_{ex}dS\nonumber\\
&=& \int_{\Omega}\sum_{i=1}^N \left\{ \nabla\tilde{\mu}_i\cdot \boldsymbol{j}_i\right\} dx-\int_{\Gamma}(-\sum_i^N\tilde{\mu}_i \gamma_i+\tilde{\mu}_e \Delta z)\mathcal{R}dS -\int_{\Gamma}\phi\frac{\partial  }{\partial t}\left(\boldsymbol{D}\cdot\boldsymbol{n} - FC_e\right)dS \nonumber\\
&&+\int_{\Gamma}(\tilde{\mu}_e-\tilde{\mu}_{ex}) j_{ex}dS +\int_{\Gamma}\tilde{\mu}_{ex} J_{ex}dS-\int_{\partial\Omega/\Gamma}\phi\frac{\partial \boldsymbol{D}\cdot\boldsymbol{n}}{\partial t}dS-\int_{\partial\Omega/\Gamma}\sum_i^N\tilde{\mu}_i \boldsymbol{j}_i\cdot\boldsymbol{n}dS\nonumber\\
&=&-\Delta+P_{E,\partial\Omega}.\nonumber
\end{eqnarray}

\noindent\textbf{Eq. \eqref{eqn:dedtoneside} derivation}
\begin{eqnarray} 
&&\frac{dE}{dt}-\mathcal{P}_{E,\partial\Omega}\nonumber\\
&=& \frac{dE_{l}}{dt} + \frac{dE_{r}}{dt} + \frac{dE_{\Gamma}}{dt}- \int_{\partial\Omega}\phi_{ref}\frac{\partial \mathbf{D}\cdot\mathbf{n}}{\partial t} dx\nonumber\\
&=&\int_{\Omega_l}\sum_{i=1}^N \nabla\tilde{\mu}_i^l\cdot \boldsymbol{j}_i^l dx -\int_{\Gamma}\sum_{i=1}^N \tilde{\mu}_i^l \boldsymbol{j}_i^l\cdot \boldsymbol{n}^l dS -\int_{\Gamma}\phi_l\frac{\partial }{\partial t}(\boldsymbol{D}^l\cdot\boldsymbol{n}^l-FC_e^l)dS + \int_{\Gamma}\tilde{\mu}_e^l(-\Delta z^l\mathcal{R}_l-J_e)\nonumber\\
&& \int_{\Omega_r}\sum_{i=1}^N \nabla\tilde{\mu}_i^r\cdot \boldsymbol{j}_i^r dx -\int_{\Gamma}\sum_{i=1}^N \tilde{\mu}_i^r \boldsymbol{j}_i^r\cdot \boldsymbol{n}^r dS -\int_{\Gamma}\phi_r\frac{\partial }{\partial t}(\boldsymbol{D}^r\cdot\boldsymbol{n}^r-FC_e^r)dS + \int_{\Gamma}\tilde{\mu}_e^r(-\Delta z^l\mathcal{R}_r+J_e)\nonumber\\
&& -\int_{\partial\Omega}(\phi-\phi_{ref})\frac{\partial \boldsymbol{D}\cdot\boldsymbol{n}}{\partial t} dS\nonumber\\
&=&\int_{\Omega_l}\sum_{i=1}^N \nabla\tilde{\mu}_i^l\cdot \boldsymbol{j}_i^l dx +\int_{\Gamma}\sum_{i=1}^N \tilde{\mu}_i^l \gamma_i\mathcal{R}^l dS -\int_{\Gamma}\phi_l\frac{\partial }{\partial t}(\boldsymbol{D}^l\cdot\boldsymbol{n}^l-FC_e^l)dS + \int_{\Gamma}\tilde{\mu}_e^l(-\Delta z^l\mathcal{R}_l-J_e)\nonumber\\
&& \int_{\Omega_r}\sum_{i=1}^N \nabla\tilde{\mu}_i^r\cdot \boldsymbol{j}_i^r dx +\int_{\Gamma}\sum_{i=1}^N \tilde{\mu}_i^r \gamma_i\mathcal{R}^r dS -\int_{\Gamma}\phi_r\frac{\partial }{\partial t}(\boldsymbol{D}^r\cdot\boldsymbol{n}^r-FC_e^r)dS + \int_{\Gamma}\tilde{\mu}_e^r(-\Delta z^l\mathcal{R}_r+J_e)\nonumber\\
&& -\int_{\partial\Omega}(\phi-\phi_{ref})\frac{\partial \boldsymbol{D}\cdot\boldsymbol{n}}{\partial t}dS \nonumber\\
&=&\int_{\Omega_l}\sum_{i=1}^N \nabla\tilde{\mu}_i^l\cdot \boldsymbol{j}_i^l dx -\int_{\Gamma}(-\sum_{i=1}^N \tilde{\mu}_i^l \gamma_i\mathcal{R}^l+\tilde{\mu}_e^l\Delta z^l)\mathcal{R}_l dS -\int_{\Gamma}\phi_l\frac{\partial }{\partial t}(\boldsymbol{D}^l\cdot\boldsymbol{n}^l-FC_e^l)dS \nonumber \\
&& +\int_{\Omega_r}\sum_{i=1}^N \nabla\tilde{\mu}_i^r\cdot \boldsymbol{j}_i^r dx -\int_{\Gamma}\left(-\sum_{i=1}^N \tilde{\mu}_i^r \gamma_i+\Delta z^r\tilde{\mu}_e^r\right)\mathcal{R}^r dS -\int_{\Gamma}\phi_r\frac{\partial }{\partial}(\boldsymbol{D}^r\cdot\boldsymbol{n}^r-FC_e^r)dS \nonumber\\
&& + \int_{\Gamma}(\tilde{\mu}_e^r
-\tilde{\mu}_e^l)J_e-\int_{\partial\Omega}(\phi-\phi_{ref})\frac{\partial \boldsymbol{D}\cdot\boldsymbol{n}}{\partial t}dS\nonumber\\
&=&  -\Delta.
\end{eqnarray}

\section{Numerical Scheme}
Here we present the detailed numerical scheme for the bidomain reaction system \eqref{bidomain system}-\eqref{bidomain system bd}.  

\begin{equation} 
\left\{
\begin{array}{l}
\frac{ (C^s_i)^{n+1} -(C^s_i)^{n} }{\Delta t} = \nabla\cdot (D_i^s (\nabla (C_i^s)^{n+1}+z_i (C_i^s)^{n+1}\nabla(\phi^s)^{n}),\\[3mm]
-\nabla\cdot(\delta^2\nabla (\phi^s)^{n+1}) =   \sum_{i=1}^N z_i (C_i^s)^{n+1},
\end{array}
\right.
\end{equation}
with the boundary conditions  
\begin{equation} 
\left\{
\begin{array}{ll}
(\boldsymbol{j}^s_i)^{n+1}\cdot\boldsymbol{n}_s =  -\gamma_i \zeta(\mathcal{R}^s)^{n},  & \mbox{on~}  \Gamma,\\
 -\nabla(\phi_s)^{n+1}\cdot\boldsymbol{n}_s = \lambda_s ((\phi^s)^{n+1}-(\phi_p^s)^{n+1}),  & \mbox{on~} \Gamma,\\    
~ \lambda_l \frac{((\phi^l)^{n+1}-(\phi_p^l)^{n+1})-  ((\phi^l)^{n}-(\phi_p^l)^{n})}{\Delta t} +\frac{\zeta }{\delta^2}\Delta z_l (\mathcal{R})^{n}_l  = g \left(\ln(\frac{(C_e^r)^{n+1}}{(C_e^l)^{n+1}})+(\phi_p^l)^{n+1}-(\phi_p^r)^{n+1}\right) ,   & \mbox{on~} \Gamma,\\    
 \lambda_r \frac{((\phi^r)^{n+1}-(\phi_p^r)^{n+1})-  ((\phi^r)^{n}-(\phi_p^r)^{n})}{\Delta t}  +\frac{\zeta  }{\delta^2}\Delta z_r (\mathcal{R})^{n}_r =-g \left(\ln(\frac{(C_e^r)^{n+1}}{(C_e^l)^{n+1}})+(\phi_p^l)^{n+1}-(\phi_p^r)^{n+1}\right), & \mbox{on~} \Gamma,\\
(\boldsymbol{j}^s)^{n+1}\cdot\boldsymbol{n} = 0, \phi^s = 0, &\mbox{on~}\partial\Omega/\Gamma,
\end{array}\right.
\end{equation} 
where  
\begin{eqnarray}
  (\mathcal{R}^l)^{n} &=&  k_{f,0}^l e^{-\Delta Z_l \beta((\phi_p^l)^n-(\phi^l)^n) }\Pi_{i=1}^N\left((C^l_i)^n\right)^{a_i} -k_{r,0}^le^{ \Delta Z_l(1-\beta)((\phi_p^l)^n-(\phi^l)^n) }((C_e^l)^n)^{-\Delta z^l}\Pi_{i=1}^N\left( (C^l_i)^n \right)^{b_i} \nonumber\\
    ( \mathcal{R}^r)^{n} &=& k_{f,0}^re^{-\Delta Z_r \beta((\phi_p^r)^{n}-(\phi^r)^n) } ((C_e^r)^n)^{\Delta z^r}\Pi_{i=1}^N\left((C^r_i)^n\right)^{a_i} -k_{r,0}^re^{ \Delta Z_r(1-\beta)((\phi_p^r)^n-(\phi^r)^n) }\Pi_{i=1}^N\left( (C^r_i)^n \right)^{b_i} .\nonumber
\end{eqnarray}
Finite volume methods \cite{xu2019numerical} are used for space discretization. 
\end{document}